\documentclass[iop]{emulateapj}
\usepackage{url}
\shorttitle{MARVELS-2b}
\shortauthors{Scott W. Fleming et al.}
\begin{document}
\submitted{Accepted in \emph{The Astronomical Journal} on 22 June 2012.}
\title{Very Low-mass Stellar and Substellar Companions to Solar-like Stars from MARVELS II: A Short-period Companion Orbiting an F Star with Evidence of a Stellar Tertiary And Significant Mutual Inclination}

\author{Scott W. Fleming\altaffilmark{1,2,3},
Jian Ge\altaffilmark{1},
Rory Barnes\altaffilmark{4},
Thomas G. Beatty\altaffilmark{5},
Justin R. Crepp\altaffilmark{6},
Nathan De Lee\altaffilmark{1,7},
Massimiliano Esposito\altaffilmark{8,9},
Bruno Femenia\altaffilmark{8,9},
Leticia Ferreira\altaffilmark{10,11},
Bruce Gary\altaffilmark{7},
B. Scott Gaudi\altaffilmark{5},
Luan Ghezzi\altaffilmark{12,11},
Jonay I. Gonz\'alez Hern\'andez\altaffilmark{8,9},
Leslie Hebb\altaffilmark{7},
Peng Jiang\altaffilmark{1},
Brian Lee\altaffilmark{1},
Ben Nelson\altaffilmark{1},
Gustavo F. Porto de Mello\altaffilmark{10,11},
Benjamin J. Shappee\altaffilmark{5},
Keivan Stassun\altaffilmark{7,13},
Todd A. Thompson\altaffilmark{5},
Benjamin M. Tofflemire\altaffilmark{4,14},
John P. Wisniewski\altaffilmark{4},
W. Michael Wood-Vasey\altaffilmark{15},
Eric Agol\altaffilmark{4},
Carlos Allende Prieto\altaffilmark{8,9},
Dmitry Bizyaev\altaffilmark{16},
Howard Brewington\altaffilmark{16},
Phillip A. Cargile\altaffilmark{7},
Louis Coban\altaffilmark{15},
Korena S. Costello\altaffilmark{15},
Luis N. da Costa\altaffilmark{12,11},
Melanie L. Good\altaffilmark{15},
Nelson Hua\altaffilmark{15},
Stephen R. Kane\altaffilmark{17},
Gary R. Lander\altaffilmark{15},
Jian Liu\altaffilmark{1},
Bo Ma\altaffilmark{1},
Suvrath Mahadevan\altaffilmark{2,3},
Marcio A. G. Maia\altaffilmark{12,11},
Elena  Malanushenko\altaffilmark{16},
Viktor Malanushenko\altaffilmark{16},
Demitri Muna\altaffilmark{18},
Duy Cuong Nguyen\altaffilmark{1,19},
Daniel Oravetz\altaffilmark{16},
Martin Paegert\altaffilmark{7},
Kaike Pan\altaffilmark{16},
Joshua Pepper\altaffilmark{7},
Rafael Rebolo\altaffilmark{8,9,20},
Eric J. Roebuck\altaffilmark{15},
Basilio X. Santiago\altaffilmark{21,11},
Donald P. Schneider\altaffilmark{2,3},
Alaina Shelden\altaffilmark{16},
Audrey Simmons\altaffilmark{16},
Thirupathi Sivarani\altaffilmark{22},
Stephanie Snedden\altaffilmark{16},
Chelsea L. M. Vincent\altaffilmark{15},
Xiaoke Wan\altaffilmark{1},
Ji Wang\altaffilmark{1},
Benjamin A. Weaver\altaffilmark{18},
Gwendolyn M. Weaver\altaffilmark{15},
Bo Zhao\altaffilmark{1}}

\email{scfleming@psu.edu}
\altaffiltext{1}{Dept. of Astronomy, University of Florida, 211 Bryant Space Science Center, Gainesville, FL, 32611-2055 USA}
\altaffiltext{2}{Department of Astronomy and Astrophysics, The Pennsylvania State University, 525 Davey Laboratory, University Park, PA 16802, USA.}
\altaffiltext{3}{Center for Exoplanets and Habitable Worlds, The Pennsylvania State University, University Park, PA 16802, USA.}
\altaffiltext{4}{Department of Astronomy, University of Washington, P.O. Box 351580, Seattle, WA 98195, USA}
\altaffiltext{5}{Department of Astronomy, The Ohio State University, 140 West 18th Avenue, Columbus, OH 43210}
\altaffiltext{6}{Department of Astronomy, California Institute of Technology, 1200 E. California Blvd., Pasadena, CA 91125, USA}
\altaffiltext{7}{Department of Physics and Astronomy, Vanderbilt University, Nashville, TN 37235, USA}
\altaffiltext{8}{Instituto de Astrof\'{i}sica de Canarias (IAC), E-38205 La Laguna, Tenerife, Spain}
\altaffiltext{9}{Departamento de Astrof\'{i}sica, Universidad de La Laguna, 38206 La Laguna, Tenerife, Spain}
\altaffiltext{10}{Universidade Federal do Rio de Janeiro, Observat\'{o}rio do Valongo, Ladeira do Pedro Ant\^{o}nio, 43, CEP: 20080-090, Rio de Janeiro, RJ, Brazil}
\altaffiltext{11}{Laborat\'{o}rio Interinstitucional de e-Astronomia, - LIneA, Rua Gal. Jos\'{e} Cristino 77, Rio de Janeiro, RJ - 20921-400, Brazil}
\altaffiltext{12}{Observat\'{o}rio Nacional, Rua General Jos\'{e} Cristino, 77, 20921-400 S\~{a}o Crist\'{o}v\~{a}o, Rio de Janeiro, RJ, Brazil}
\altaffiltext{13}{Department of Physics, Fisk University, 1000 17th Ave. N., Nashville, TN 37208, USA}
\altaffiltext{14}{Astronomy Department, University of Wisconsin-Madison, 475 N Charter St, Madison, WI 53706, USA}
\altaffiltext{15}{Department of Physics and Astronomy, University of Pittsburgh, Pittsburgh, PA 15260, USA}
\altaffiltext{16}{Apache Point Observatory, P.O. Box 59, Sunspot, NM 88349-0059, USA}
\altaffiltext{17}{NASA Exoplanet Science Institute, Caltech, MS 100-22, 770 South Wilson Avenue, Pasadena, CA 91125, USA}
\altaffiltext{18}{Center for Cosmology and Particle Physics, New York University, New York, NY 10003 USA}
\altaffiltext{19}{Department of Physics and Astronomy, University of Rochester, Rochester, NY 14627, USA}
\altaffiltext{20}{Consejo Superior de Investigaciones Cient\'{i}ficas, Spain}
\altaffiltext{21}{Instituto de F\'{i}sica, UFRGS, Caixa Postal 15051, Porto Alegre, RS - 91501-970, Brazil}
\altaffiltext{22}{Indian institute of Astrophysics, 2nd block, Koramangala, Bangalore 560034, India}

\begin{abstract}
We report the discovery via radial velocity measurements of a short-period ($P = 2.430420 \pm 0.000006$ days) companion to the F-type main sequence star TYC 2930-00872-1.  A long-term trend in the radial velocity data also suggests the presence of a tertiary stellar companion with $P > 2000$ days.  High-resolution spectroscopy of the host star yields $T_{\rm eff} = 6427 \pm 33~{\rm K}$, $\log{g}=4.52 \pm 0.14$, and [Fe/H]=$-0.04\pm 0.05$.  These parameters, combined with the broad-band spectral energy distribution and a parallax, allow us to infer a mass and radius of the host star of $M_1=1.21 \pm 0.08~\rm{M_\odot}$ and $R_1=1.09_{-0.13}^{+0.15}~\rm{R_\odot}$.  The minimum mass of the inner companion is below the hydrogen burning limit, however the true mass is likely to be substantially higher.  We are able to exclude transits of the inner companion with high confidence.  Further, the host star spectrum exhibits a clear signature of Ca H and K core emission indicating stellar activity, but a lack of photometric variability and small $v\sin I$ suggest the primary's spin axis is oriented in a pole-on configuration.  The rotational period of the primary estimated through an activity-rotation relation matches the orbital period of the inner companion to within $1.5\,\sigma$, suggesting that the primary and inner companion are tidally locked.  If the inner companion's orbital angular momentum vector is aligned with the stellar spin axis as expected through tidal evolution, then it has a stellar mass of $\sim 0.3-0.4~\rm{M_\odot}$.  Direct imaging limits the existence of stellar companions to projected separations $< 30$ AU.  No set of spectral lines and no significant flux contribution to the spectral energy distribution from either companion are detected, which places individual upper mass limits of $M_{\left\{2,3\right\}} \lesssim 1.0 ~ \rm{M_{\odot}}$, provided they are not stellar remnants. If the tertiary is not a stellar remnant, then it likely has a mass of $\sim 0.5-0.6~\rm{M_\odot}$, and its orbit is likely significantly inclined from that of the secondary, suggesting that the Kozai-Lidov mechanism may have driven the dynamical evolution of this system.
\end{abstract}

\section{Introduction}
Exoplanet surveys have contributed to a wide range of ancillary astrophysical disciplines during the last two decades, including studies of variable stars, binary stars and brown dwarf (BD) companions.  During the course of operation, these surveys detect a large variety of stellar binaries that can be used to study stellar structure, atmospheres and formation mechanisms.  One example of the latter is a study of the multiplicity of close binaries, e.g., the fraction of close binaries that are in triple or higher-order systems.  Indeed, triple systems are not uncommon amongst short-period binaries; 9 out of 16 binaries with $P < 100$ days in the volume-limited sample of \citet{rag2010} are members of triple systems.  Shorter-period binaries have an even greater probability of being in a multiple-star system \citep[$\sim 80$\% for $P < 7$ days vs. $\sim 40$\% for $P > 7$ days,][]{tok2006}.

The orbital elements of such binaries, including the mutual inclinations of the companions' orbital angular momentum vectors, are fossil records of their formation process, and provide critical constraints to binary star formation models \citep{ste2002}.  Comparison of the orbital and physical properties between different binary hierarchies also provides insight into binary star formation theory \citep{tok2008}.  In fact, the dynamical evolution of these systems may be dominated by dynamical interactions between the inner and outer companions via a combination of the Kozai-Lidov mechanism \citep{koz1962,lid1962} and tidal forces, which drive the inner companion to shorter orbital separations until it circularizes with some period $P \lesssim 10$ days, beyond which tidal forces are ineffective \citep{fab2007}.

In this paper, we present the discovery of a companion with a substellar minimum mass orbiting the bright ($V = 9.8$) F-type star TYC 2930-00872-1 \citep[][hereafter TYC 2930]{hog2000}, with an orbital period of $P = 2.430420 \pm 0.000006$ days.  This discovery is part of a series of papers dedicated to analyses of individual low-mass companions in anticipation of a global analysis of the MARVELS sample at the conclusion of the survey \citep[e.g.,][]{lee2011,wis2012}, therefore, TYC 2930 is also designated ``MARVELS-2'' as an internal reference within this series.  The a priori transit probability of the inner companion is $\sim$13\% with an expected central transit depth of ${\sim}0.9 \, \pm \, 0.25$\% for a $1 ~ \rm{R_{Jup}}$ companion radius, although no transits are detected.  An additional, long-term trend in the RV data is detected from a stellar tertiary in the system.  A detailed analysis of the combined radial velocity, spectroscopic and photometric data suggests the inner companion is oriented towards a pole-on configuration, and is more likely an M dwarf with a mass $\sim 0.3-0.4~\rm{M_{\odot}}$, while the tertiary is likely to be less inclined.  In such a scenario, the mutual inclination between the secondary and tertiary is likely to be significant, which would make this an excellent example of a system whose dynamical history was driven via the Kozai-Lidov mechanism.

The paper is organized such that \S \ref{specsec} describes the spectroscopic observations and their data processing, \S \ref{photobs} describes the archival and observed photometry for the system, \S \ref{starchar} describes the characterization of the host star's properties, including mass, radius, effective temperature, surface gravity, metallicity, stellar activity and rotation rate, \S \ref{orbitfit} describes our determination of the orbital parameters from fitting the measured RVs, \S \ref{imaging} describes both Lucky Imaging and adaptive optics imaging to search for any wide companions to TYC 2930, \S \ref{relphot} describes our search for photometric variability and any potential transits of the inner companion, \S \ref{geomtides} discusses the tidal evolution of the inner companion, \S \ref{massdistsection} describes the posterior distribution of the true masses for both the secondary and tertiary given the results from the previous sections, and finally, \S \ref{kozaisection} investigates the possible dynamical history of the system via the Kozai-Lidov mechanism.

\section{Description of Observations}
\subsection{Spectroscopic Observations}
\label{specsec}
MARVELS \citep[Multi-object APO Radial Velocity Exoplanets Large-area Survey, ][]{ge2008} is part of the Sloan Digital Sky Survey III \citep[SDSS-III, ][]{eis2011}.  The instrument uses dispersed fixed-delay interferometry \citep[DFDI;][]{gee2002,ge2002,ers2002,ers2003,van2011} on the 2.5m SDSS telescope \citep{gun2006} at Apache Point Observatory (APO) to measure precision radial velocities of 60 stars simultaneously.  Both beams of the interferometer are imaged onto the detector for a given star, for a total of 120 spectra, producing two simultaneous RV measurements for each star from beams that travel through a slightly different instrument path.  The survey began in the Fall of 2008 and will ultimately target several thousand stars between $7.6 < V < 12$, with a baseline goal of $< 30 ~ \rm{m~s^{-1}}$ precision for the faintest stars.  Each star is observed $\sim$20-30 times over a typical baseline of 1.75 years.  In addition to exoplanets, the survey will conduct studies of stellar atmospheres, binary stars, and rare companions such as BDs and very low mass (VLM; $M \lesssim 150 ~ \rm{M_{Jup}}$) stars at short orbital periods.

TYC 2930 was observed a total of 33 times over a baseline of 707 days.  The data were processed by the MARVELS pipeline following the steps described in \citet{lee2011}.  The resultant RV measurements from both interferometer output beams were combined via a weighted average after they were found to agree to within the measurement errors.  The formal mean RV precision was $23 ~ \rm{m~s^{-1}}$.  Following \citet{fle2010}, the RV uncertainties for this star were further scaled by a ``quality factor'' $QF = 6.69$, a first-order correction used to partially account for residual systematic errors.  For each of the other 118 spectra in this field, an individual $QF$ is calculated as the RV root-mean-square (rms) about the mean, divided by the median formal RV uncertainty for that star.  On average, most of the MARVELS targets in a given field should be RV-stable at the level of tens of $\rm{m~s^{-1}}$, and therefore should have $QF \sim 1$.  Since the average $QF$ across the plate is significantly larger, we treat that as one measurement of the residual uncertainties from the pipeline-produced RVs.  The dates and RVs from the MARVELS observations are presented in Table \ref{marvrvs}.

Additional RV observations were conducted using the Spettrografo Alta Risoluzione Galileo (SARG) spectrograph \citep{gra2001} on the 3.58m Telescopio Nazionale Galileo (TNG) telescope.  The data were obtained using the yellow grism with a slit of $0.8{\arcsec}{\times}5.3{\arcsec}$ on-sky that produces a resolving power of $R = 57000$ over the wavelength range $462 < \lambda < 792$ nm.  A total of 20 observations were taken, spanning $\sim$408 days, using an iodine cell that serves as an RV calibration source.  The average signal-to-noise ratio per resolution element, averaged across the central 200 pixels of all the orders, ranges from 150-290.  An additional observation was taken without the iodine cell to be used as a stellar template and to derive stellar parameters.  The signal-to-noise ratio per resolution element of the template spectrum is ${\sim}400$ at 607 nm.

The SARG data are processed using the standard IRAF Echelle reduction packages.  Frames are trimmed, bias subtracted, flat-field corrected, aperture-traced and extracted.  ThAr lines are used to calibrate the wavelength solution.  The RVs are measured using the iodine cell technique \citep{mar1992}.  The 21 SARG orders that have sufficiently strong iodine lines lie in the wavelength range $504 < \lambda < 611$ nm.  Each order is subdivided into 10 sections from which an RV is measured.  The resulting 210 RV measurements are then 2-$\sigma$ clipped using three iterations.  The remaining $N$ RV measurements are averaged to produce a single RV measurement.  The RV uncertainty is given by $\sigma_{\rm{RV}} = \sigma ~ \it{N}^{\rm{-1/2}}$, where $\sigma$ is the dispersion of the points after the 2-$\sigma$ clipping.  Table \ref{tngrvs} contains the dates and RVs from the SARG observations.

A high signal-to-noise ratio spectrum of TYC 2930 was obtained with the ARC Echelle Spectrograph \citep[ARCES;][]{wan2003} on the APO 3.5m telescope for the purposes of stellar characterization.  The spectrograph delivers $R \sim$31500 spectra spanning a wavelength range $320 < \lambda < 1000 ~$ nm on a single 2048x2048 SITe CCD.  The spectra were reduced using an IRAF script that corrects for bias and dark current subtraction, cosmic rays and bad pixels.  Flat-Fielding is performed using a combination of quartz lamp exposures with and without a blue filter, while a ThAr lamp is used for wavelength calibration.  A single integration of 900 seconds was taken, yielding a spectrum with a signal-to-noise ratio of $\sim$220 per resolution element at 607 nm.

Long-term, queue-scheduled \citep{she2007}, RV monitoring of the TYC 2930 system has been initiated using the High Resolution Spectrograph \citep[HRS,][]{tul1998} on the Hobby-Eberly Telescope \citep[HET,][]{ram1998} to further characterize the orbit of the suspected long-period companion.  These RVs are expected to be presented in a separate paper at the conclusion of that project.

\subsection{Photometry Observations}
\label{photobs}
Photometry of TYC 2930 was performed using the Hereford Arizona Observatory (HAO), a private observatory in southern Arizona (observatory code G95 in the IAU Minor Planet Center).  Observations were taken in Johnson $B$ and $V$ filters using a Meade 14-inch LX200GPS telescope and a 2184x1472 pixel SBIG ST-10XME CCD.  Landolt standard stars \citep{lan2007,lan2009} were observed in the Kapteyn Selected Area 98 (SA 98) for calibration.  A photometric precision of 0.023 mag was obtained in $B$ and 0.018 in $V$.  The measured fluxes and uncertainties are presented in Table \ref{stellarprops}.

We obtained relative photometric time series from several ground-based telescopes (SuperWASP, Allegheny Observatory, KELT-North) to search for transits and examine the photometric stability of the primary star.  We briefly describe each of these data sets in turn.  The SuperWASP instruments measure fluxes of millions of stars via wide-angle images of the night sky using a broad-band filter that covers 400-700 nm, and are described in \citet{pol2006}.  For TYC 2930, a total of 2204 observations from 2006 and 1309 observations from 2007 are extracted from the SuperWASP public archive\footnote{\url{http://www.wasp.le.ac.uk/public/}}.

We obtained photometric observations on 7 nights in February and March of 2011 using the Keeler 16-inch Meade RCS-400 telescope at Allegheny Observatory.  The CCD detector is a $3060{\times}2040$ pixel SBIG KAF-6303/LE with a $0.57\arcsec$ per pixel scale, and all observations were taken through a Johnson-Cousins R filter.  Typical seeing was $2.5\arcsec$ with integration times ranging from 20-30 seconds.  The images were processed using standard bias, dark, and flat-field calibration images taken on the same nights.  Astrometric solutions were computed based on the positions of stars in the $20{\arcmin}{\times}30{\arcmin}$ field of view from the 2MASS Point-Source Catalog \citep{skr2006}.  After image calibration, we performed circular aperture photometry with a 10-pixel radius ($5.7\arcsec$ on sky) and estimated the local sky background from a 15-20 pixel annulus around each star.  Relative photometry was determined by comparing the measured flux from TYC 2930 with two nearby stars.

We also extracted photometric time series data of TYC 2930 obtained by the Kilodegree Extremely Little Telescope (KELT) North transit survey \citep{pep2007,siv2009}. KELT uses a red-pass filter with a 50\% transmission point at 490 nm, which, when folded with the CCD response, yields an effective bandpass similar to R, but broader. The standard KELT data reduction procedure uses the ISIS image subtraction package \citep{ala1998}, combined with point-spread fitting photometry using DAOPHOT \citep{ste1987}. 

In the case of TYC 2930, the standard KELT data reduction procedure yielded an unusable lightcurve due to the presence of the nearby bright star HD 42903, which was partially saturated in the KELT images.  We correct the systematics by performing simple aperture photometry on both TYC 2930 and HD 42903 using the subtracted images. We used two apertures centered on HD 42903, and one aperture centered on TYC 2930. We sized the apertures around HD 42903 such that they formed an annulus that included the systematic artifacts. The single aperture around TYC 2930, in the middle of the artifact, had the same diameter as the width of the annulus around HD 42903. By subtracting the summed flux in the aperture around TYC 2930 from the annulus around HD 42903, we are left with the average negative flux value in the artifact for each of the subtracted images. We then used this average value to correct the results from the aperture around TYC 2930.  This procedure was tested using known variable stars and on stars with similar brightness ratios and angular separations to confirm that accurate results were obtained and no intrinsic variations were suppressed.

\section{Stellar Characterization}
\label{starchar}
\subsection{Stellar Parameters}
TYC 2930 (HIP 29714) is a bright F-type star located 6.16$\arcmin$ from the center of the open cluster NGC 2192.  The Hipparcos parallax measurement \citep{van2007} places the star at a distance of $d = 140 \pm 29.5$ pc.  The RPM-J \citep{col2007} value of 2.14 and $\left(J-H\right)$ color of 0.23 are consistent with a main sequence star.  We further characterize the host star's properties using the SARG template spectrum and the ARCES spectrum by measuring equivalent widths (EWs) of \ion{Fe}{1} and \ion{Fe}{2} lines.  We utilize two independent pipelines that derive stellar atmospheric parameters based on the \ion{Fe}{1} and \ion{Fe}{2} excitation and ionization equilibria.  We refer to these different pipelines as the ``IAC'' (Instituto de Astrof\'{i}sica de Canarias) and ``BPG'' (Brazilian Participation Group) pipelines, which are described in detail by \citet{wis2012}.  We apply both these analyses to the SARG and ARCES spectra, and find spectroscopic parameters that are consistent across both groups and both instruments.  We summarize the individual $T_{\rm eff}$, $\log{(g)}$, [Fe/H], and ${\xi}_{t}$ in Table \ref{bpgiacsp}.

A final, mean value for each parameter is calculated following \citet{wis2012}, yielding $T_{\rm eff} = 6427 \pm 33 $ K, $\log{(g)} = 4.52 \pm 0.14$, $\rm{[Fe/H]} = -0.04 \pm 0.05$ and ${\xi}_{t} = 1.40 \pm 0.05 ~ \rm{km ~ s^{-1}}$.  We note that while there can be correlations between the measured stellar parameters, we treat their uncertainties as independent and Gaussian-distributed in this analysis.  Estimates of the primary's mass and radius are determined using a Markov chain Monte Carlo (MCMC) analysis applied to  the empirical relationship of \citet{tor2010}, the Hipparcos parallax, and the stellar parameters described above.  The uncertainties for the mass and radius include the correlations of the best-fit coefficients from \citet{tor2010} and the reported scatter in that relation ($\sigma_{\log{m}} = 0.027$ and $\sigma_{\log{r}} = 0.014$).  The radius of the primary is $R = 1.09 ^{+0.15} _{-0.13} ~ \rm{R_\odot}$ and the mass is $M = 1.21 \pm 0.08 ~ \rm{M_\odot}$.  All of the stellar parameters are summarized in Table \ref{stellarprops}.

\subsection{Rotation Rate, SED Fitting, and Stellar Activity}
We measure the stellar rotational velocity $v\sin I$ from the SARG template spectrum.  Note that we utilize a notation that distinguishes $I$ (the angle between our line-of-sight and the stellar rotation axis) from $i$ (the angle between our line-of-sight and a companion's orbital angular momentum vector).  We use an interpolated Kurucz model spectrum \citep{kur1993} using the spectroscopically determined $T_{\rm eff}$, $\log{(g)}$ and $\rm{[Fe/H]}$, convolved to the instrumental profile (FWHM of $5.3 ~ \rm{km ~ s^{-1}}$).  Testing showed that macroturbulence (${\zeta}_{t}$) had only a marginal affect on the final result, so we adopt values ranging from 2-5 $\rm{km ~ s^{-1}}$.  The model spectra are broadened with a Gray's profile over a range of $v\sin I$ from 0-10 $\rm{km ~ s^{-1}}$.  These models are then compared via a $\chi^2$ analysis with the observed spectrum, which yields $v\sin I = 3.8 \left(+1.9, -2.8\right) ~ \rm{km ~ s^{-1}}$, effectively placing an upper limit of $v\sin I \la 6 ~ \rm{km ~ s^{-1}}$.

\begin{figure}
\includegraphics[angle=270,scale=0.7]{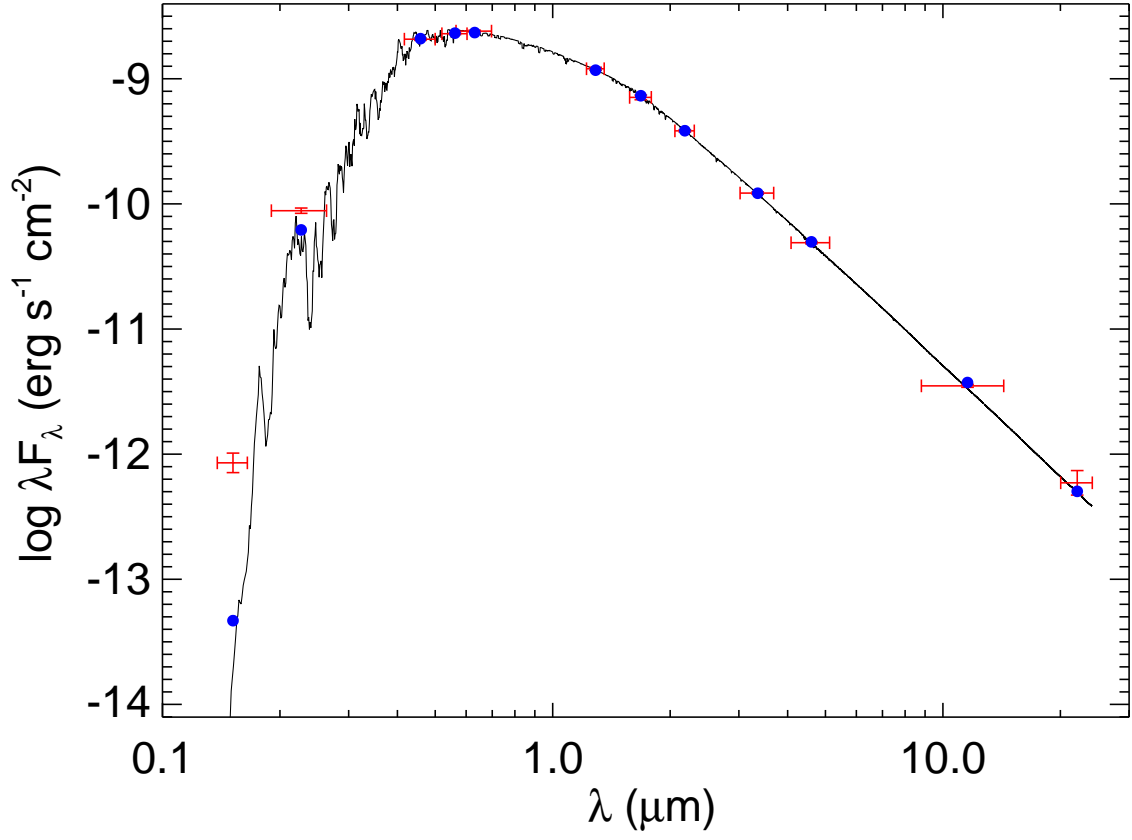}
\caption{NextGen model (solid line) compared to the observed broadband fluxes of the host star.  Blue points represent the expected fluxes in each band based on the model, red horizontal bars are the approximate bandpass widths, and red vertical bars are the flux uncertainties.  The $T_{\rm eff}$, $\log{(g)}$ and [Fe/H] are fixed at the spectroscopically-determined values, while $A_{\rm{V}}$ is allowed to float.  No evidence of IR excess is detected, while there is potentially some GALEX FUV excess indicating elevated levels of stellar activity.\label{sedfit}}
\end{figure}

\begin{figure}
\includegraphics[angle=90,scale=0.7]{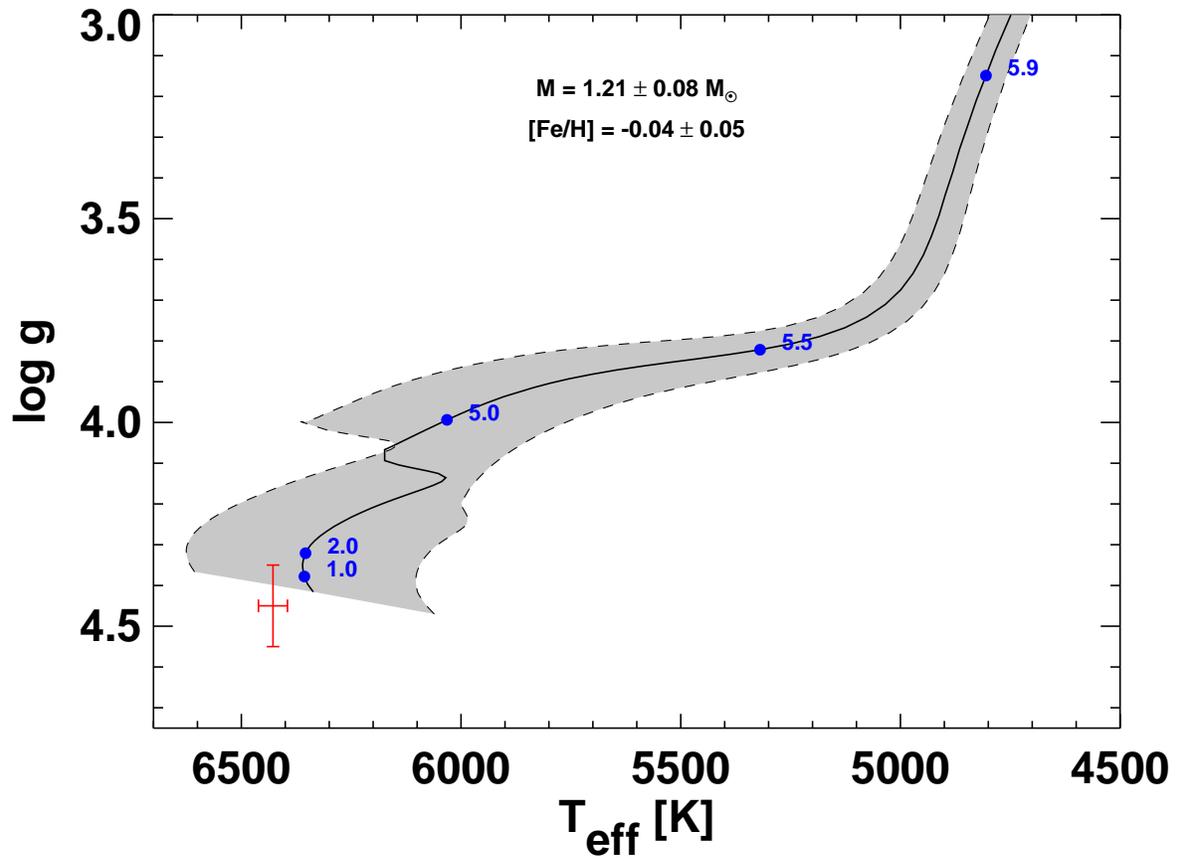}
\caption{HR diagram based on Yonsei-Yale stellar evolution models \citep{dem2004}.  The solid track is for the best-fit stellar parameters, while the two dashed tracks represent the 1-$\sigma$ uncertainties.  The blue dots represent star ages in Gyr.  TYC 2930 (red point) appears to lie on the main sequence.\label{hrdiagram}}
\end{figure}

We construct an SED using fluxes from GALEX \citep{mar2005}, the HAO observations, the 2MASS \citep{skr2006} Point Source Catalog, and the four WISE \citep{wri2010} bands.  NextGen models from \citet{hau1999} are used to construct theoretical SEDs by fixing $T_{\rm eff}$, $\log{(g)}$ and [Fe/H] at the spectroscopic values, while the extinction $A_{\rm{V}}$ is constrained to a maximum value of $A_{\rm{V}} = 0.6$ based on the reddening maps of \citet{sch1998} for galactic coordinates $\left(l,b\right) = \left(173.365948 ^{\circ}, 10.729936 ^{\circ}\right)$.  Fig.\ \ref{sedfit} shows the best-fit model, which has a $\chi^2/{\rm dof}=1.2$, $A_{\rm{V}} = 0.33 \pm 0.06$, no evidence for IR excess, and some excess in the GALEX FUV band.  Fig.\ \ref{hrdiagram} places the star on an HR diagram based on Yonsei-Yale stellar models \citep{dem2004}, indicating that TYC 2930 is consistent with an F-type dwarf with an age $t < 2$ Gyr.

\begin{figure}
\includegraphics[scale=0.6]{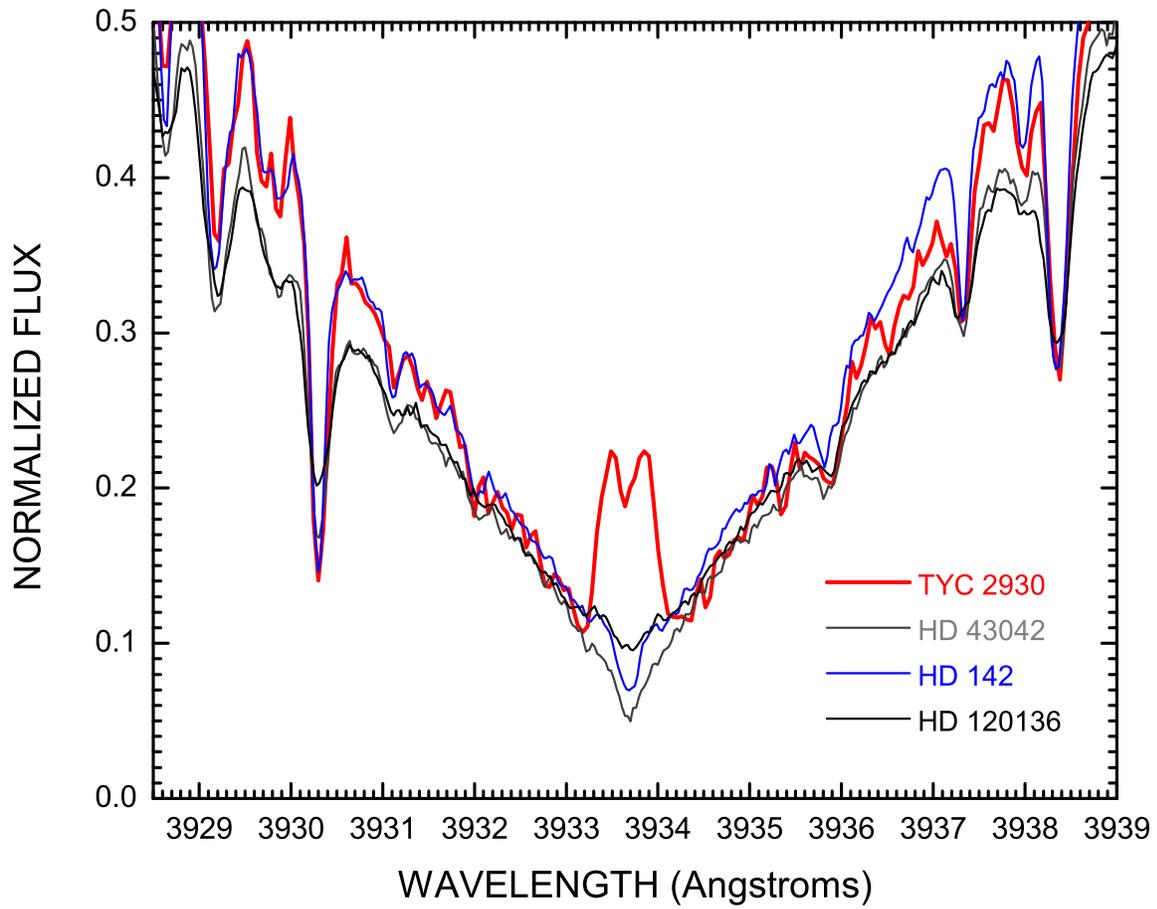}
\caption{Ca II K line of TYC 2930 compared to standard stars.  TYC 2930 has clear core emission indicating that the host star is active compared to other stars with similar stellar parameters.\label{cakcomp}}
\end{figure}

To further explore the FUV excess, Fig.\ \ref{cakcomp} compares the ARCES spectrum of TYC 2930, centered on the Ca II K line at 393.37 nm, with archival Fibre-fed Extended Range Optical Spectrograph \citep[FEROS,][]{kau1999} spectra of the standard stars HD 43042, HD 142 and HD 120136 from \citet{ghe2010}.  FEROS is an $R \sim 48000$ spectrograph with a wavelength range of $350 < \lambda < 920$ nm and high throughput ($\sim 20$\%).  There is clear Ca K core emission from TYC 2930 indicating significant chromospheric activity.  We measure the $S$ index \citep{vau1978,vau1980,dun1991} from the APO spectrum and convert to $R^{\prime}_{HK}$, finding a $\log{R^{\prime}_{HK}} = -4.44 \pm 0.05$.

\section{Orbital Analysis and Companion Minimum Mass}
\label{orbitfit}
\subsection{Radial Velocity Fitting}
\label{rvfit}
The MARVELS RVs show evidence of a long-term, positive linear trend indicating a possible tertiary object.  The TNG data also showed evidence of a long-term trend, but with a negative slope.  Fitting the MARVELS+SARG data with a single-companion model combined with non-Keplerian trends (linear, parabolic, cubic) yielded residuals with significant systematics, suggesting a two-companion Keplerian model is required.

\begin{figure}
\includegraphics[scale=0.95]{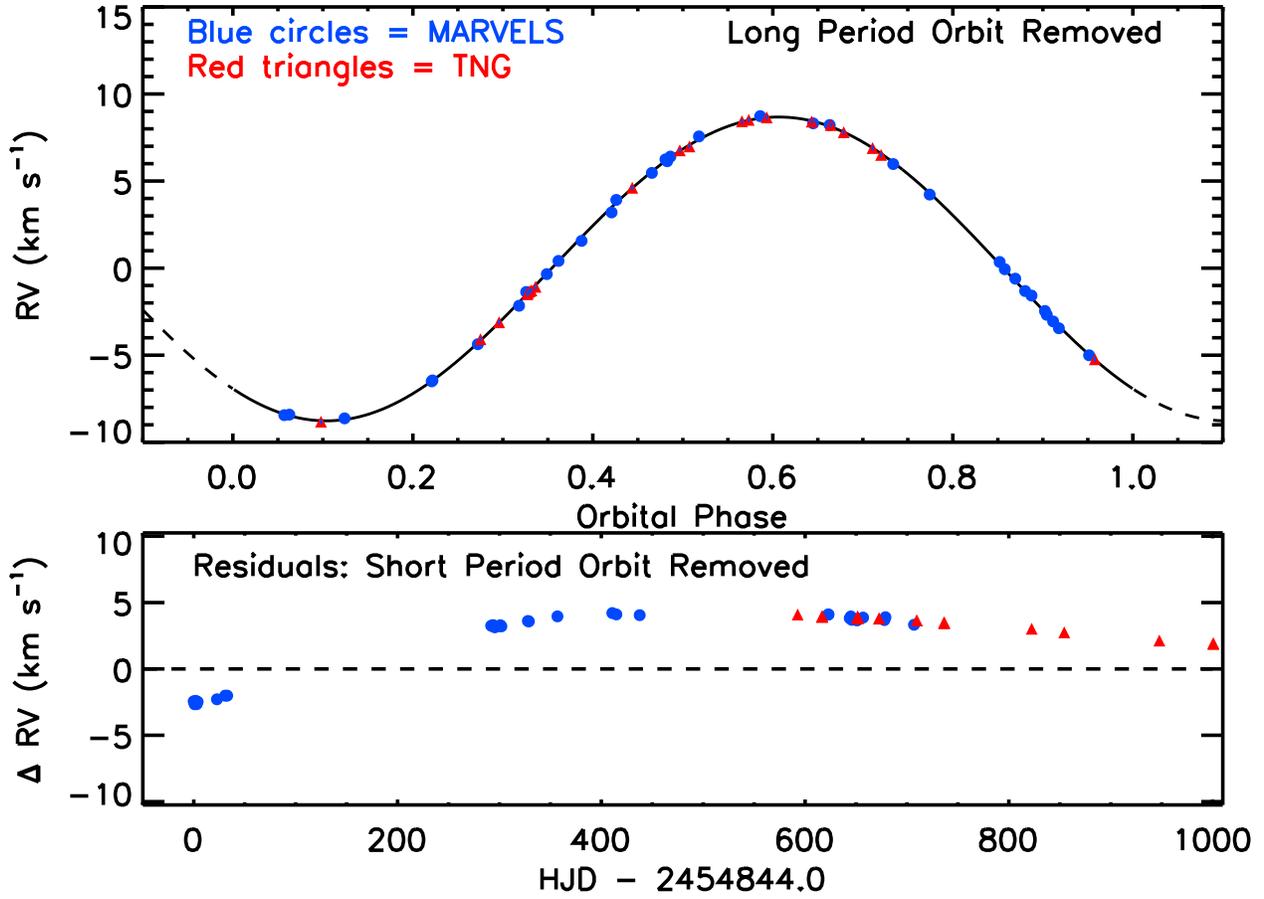}
\caption{Phase-folded MARVELS (blue) and SARG (red) RVs of MARVELS-2b.  The orbit of the long-period companion and the systemic velocity ($\gamma_{0,\rm{inst}} = -6.642 ~ \rm{km ~ s^{-1}}$) have been removed in the top panel.  The bottom panel shows the residual RVs after removing the short-period companion's orbit and systemic velocity.  The RV uncertainties are not visible at this scale.\label{rvshortper}}
\end{figure}

\begin{figure}
\includegraphics[scale=0.95]{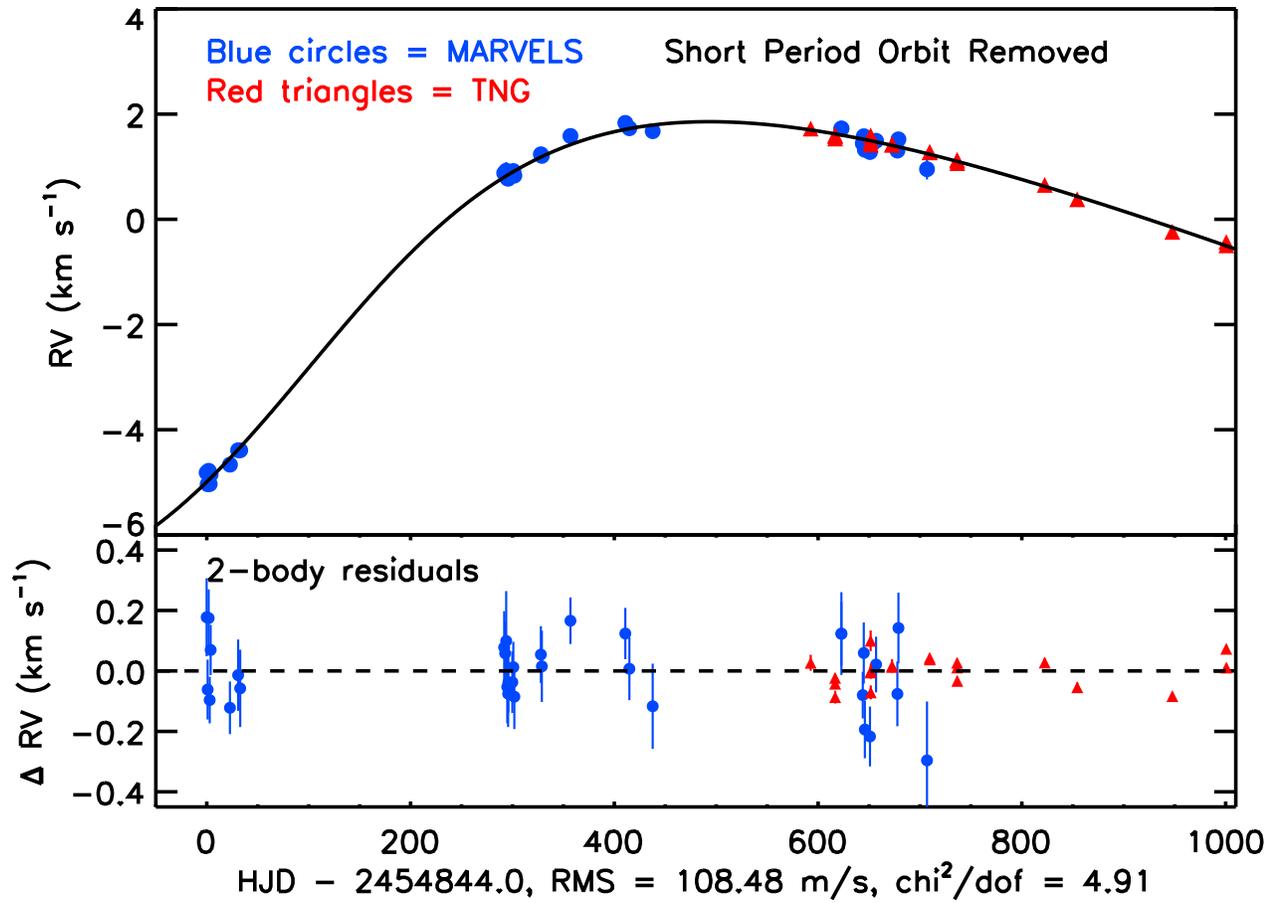}
\caption{Unfolded MARVELS (blue) and SARG (red) RVs after removing the short-period companion's orbit and the systemic velocity ($\gamma_{0,\rm{inst}} = -6.642 ~ \rm{km ~ s^{-1}}$).  The best-fit model of the outer companion is overplotted as the black line.  The residuals of the combined, two-companion model are shown in the bottom panel.\label{rvlongper}}
\end{figure}

The combined RVs were initially fit using the RVLIN package \citep{wri2009} for the purpose of obtaining initial values of the orbital parameters.  Uncertainties are calculated later using MCMC analysis.  The initial best-fit orbital period for the inner companion is $P_2 = 2.430420$ days, with a semi-amplitude $K_2 = 8724 ~ \rm{m ~ s^{-1}}$ and an eccentricity that is consistent with a circular orbit.  Fig.\ \ref{rvshortper} shows the MARVELS (blue) and SARG (red) RVs phase-folded on the best-fit orbital solution for this inner companion after removing the effects of the longer period orbit.  The bottom panel plots the residual RVs after removing the shorter period orbit.  The unfolded RVs from MARVELS (blue) and SARG (red) are shown in Fig.\ \ref{rvlongper}, where the shorter period orbit has been removed and the best-fit model of the longer period orbit is plotted as the solid line.  The residuals of the combined, two-companion solution are shown in the bottom panel.

To derive final orbital parameters and associated uncertainties, we perform an MCMC analysis closely following the methods of \citet{for2006}. For review, our goal is to estimate the uncertainties in our set of model parameters, $\theta=\{P_2$, $P_3$, $K_2$, $K_3$, $e_2$, $e_3$, $\omega_2$, $\omega_3$, $T_{P2}$, $T_{P3}$, $\gamma_{\rm{off}}$, $\gamma_{\rm{0,inst}}\}$, where $P$ is the orbital period, $K$ is the RV semiamplitude, $e$ is the orbital eccentricity, $\omega$ is the argument of periastron, $T_p$ is the epoch of periastron, $\gamma_{\rm{off}}$ is the offset between the two sets of instruments, $\gamma_{\rm{0,inst}}$ is the (instrumental) systemic velocity, and the subscripts $j = \left\{2,3\right\}$ refer to the shorter period and longer period companions, respectively.  We sample the posterior probability distribution given by Bayes' theorem, where specifics on the priors and likelihood function can be found in Section 3 of \citet{zak2011}. To help accelerate convergence, we use additional combinations of parameters identified in Section 4 of \citet{for2006}.  We do not attempt to place constraints on stellar jitter in our model.

We test for non-convergence by monitoring the Gelman-Rubin statistic \citep{gel2003}, verifying it is less than 1.02 for each of the parameters, and that chains have been allowed to run long enough to enter these regions of parameter space at least 100 times.  The orbital parameters and 1-$\sigma$-equivalent confidence levels for the inner companion are given in Table \ref{orbprops}.  The outer companion's orbital parameters are not well-determined, since the orbital period is longer than the baseline of the measurements, but we place a lower limit of $P_3 \gtrsim 2000$ days.  The RV semiamplitude and eccentricity of the outer companion are positively correlated with the best-fit orbital period, which must be accounted for when constraining the outer companion's properties.

To place the RVs on an absolute scale, SARG spectra from 620-800 nm are cross-correlated with a high resolution solar spectrum.  To partially account for temporal variation in the slit illumination and wavelength solution, a correction is applied via cross-correlation of the telluric lines near 690 nm with a numerical mask.  The telluric line locations are taken from \citet{gri1973}; the corrections are typically a few hundred $\rm{m ~ s^{-1}}$.  After removing the barycentric velocity and orbital motion of the companions based on the MCMC parameters, we find a median absolute RV of $\gamma_{0} = 35.751 \pm 0.285 ~ \rm{km ~ s^{-1}}$, where the uncertainty is taken as the rms about the median.

The $\left\{U, V, W\right\}$ space velocities can then be calculated using the absolute RV along with the parallax and proper motion measurements from Hipparcos \citep{van2007}.  We find $\left\{U, V, W\right\} = \left\{ -31.2 \pm 0.7, -7.4 \pm 3.1, 3.0 \pm 1.6\right\} ~ \rm{km ~ s^{-1}}$, where $U$ is pointing towards the Galactic center.  From the classification scheme of \citet{ben2003}, TYC 2930 is almost certainly a member of the thin disk, as its relative probability of being a thick disk member is just $0.71 \pm 0.02$\%.

\subsection{Mass Functions of the Secondary and Tertiary}
Using the MCMC chain from the joint RV fit, we can derive the mass functions $M_j$ of companion $j=2,3$,
\begin{equation}
{\cal M}_j \equiv \frac{(M_j \sin i_j)^3}{(M_1+M_j)^2} = K_j^3 (1-e_j^2)^{3/2} \frac{P_j}{2\pi G}
\label{eqn:mf}
\end{equation}
The mass functions are the only properties of the companions that we can derive that are independent of the properties of the primary.  For the secondary, we find,
\begin{equation}
{\cal M}_2 = (1.6711 \pm 0.0050) \times 10^{-4} \rm{M_\odot},
\label{eqn:mf2}
\end{equation}
where the uncertainty is essentially dominated by the uncertainty in $K_2$, such that $\sigma_{{\cal M}_2}/{\cal M}_2 \sim  3(\sigma_{K_2}/{K_2}) = 3\times 0.1\% \sim  0.3\%$.  

For the tertiary, the uncertainty in the mass function is much larger, because of the incomplete
phase coverage of the radial velocity curve (Figs.\ \ref{rvshortper} and \ref{rvlongper}).  In particular, there is a broad tail toward high mass functions.  We therefore quote the median
and 68\% confidence interval, 
\begin{equation}
{\cal M}_3 = 2.55_{-1.22}^{+5.50} \times 10^{-2} \rm{M_\odot}.
\label{eqn:mf3}
\end{equation}

\subsection{Minimum Mass and Mass Ratio}
To determine the mass or mass ratio of the secondary and tertiary, we must estimate the mass of the primary, as well as the inclination of the secondary and tertiary.  To estimate the mass and radius of the primary, we use an MCMC chain where, for each link in the MCMC chain from the joint RV fit, we draw a value of $T_{\rm eff}$, $\log g$, and $\rm{[Fe/H]}$ for the primary from Gaussian distributions, with means and dispersions given in Table \ref{stellarprops}. We then use the \citet{tor2010} relations to estimate the mass $M_1$ and radius $R_1$ of the primary, including the intrinsic scatter in these relations.

The minimum mass (i.e., $M_2$ if $\sin i_2=1$) and minimum mass ratio of the secondary are:
\begin{eqnarray}
{M_{2,\rm min}= 68.1 \pm 3.0 ~ \rm{M_{Jup}} = 0.0650 \pm 0.0029 ~ \rm{M_\odot},} \\
{q_2 = 0.0535 \pm 0.0012}
\label{eqn:m2min}
\end{eqnarray}

The uncertainties in these estimates are almost entirely explained by the uncertainties in the mass of the primary: $\sigma_{M_2}/M_2 \sim (2/3)(\sigma_{M_1}/M_1) = (2/3)\times 6.7\% \sim 4.5\%$, almost exactly the uncertainty in $M_{2,\rm min}$ above ($4.4\%$), and $\sigma_q/q \sim (1/3)(\sigma_{M_1}/M_1) \sim (1/3)\times 6.7\% \sim 2.3\%$, the uncertainty in $q$ ($2.3\%$).  As we show in \S \ref{transearch}, an edge-on orbit for the secondary is excluded from the lack of transits for reasonable assumptions about its radius.

The minimum mass and mass ratio of the tertiary are much more poorly constrained due to the incomplete phase
coverage of the orbit.  We find median and 68\% confidence intervals of,
\begin{eqnarray}
{M_{3,\rm min}= 426_{-98}^{+261} ~ \rm{M_{Jup}} = 0.407_{-0.093}^{+0.249} ~ \rm{M_\odot},} \\
{q_3= 0.334_{-0.0761}^{+0.205}}
\label{eqn:m3min}
\end{eqnarray}
The uncertainties in these quantities contain significant contributions from both the uncertainty in the host star mass and the tertiary mass function.

\section{Imaging}
\label{imaging}
Lucky Imaging \citep{fri1978} was performed in Oct. 2010 and Oct. 2011 using FastCam \citep{osc2008} on the 1.5 m TCS telescope at Observatorio del Teide in Spain to search for companions at large separations from the primary star.  Lucky Imaging consists of taking observations at very high cadence to achieve nearly-diffraction-limited images from a subsample of the total.  During the Oct. 2011 observations, the CCD gain was adjusted, and therefore that night's data are analyzed as two different image sets.  For the Oct. 2010 data, a total of 140000 frames comprised of 50 ms integrations were obtained in the $I$ band spanning $21{\arcsec}{\times}21{\arcsec}$ on sky.  For the Oct. 2011 observations, a total of 31000 frames comprised of 50 ms integrations were obtained in the low-gain setting, and 100000 frames comprised of 40 ms integrations were obtained in the higher-gain setting.  Image selection is applied using a variety of selection thresholds (best $X$\%) based on the brightest pixel (BP) method, making sure that non-speckle features are avoided.

\begin{figure}
\includegraphics[scale=0.8]{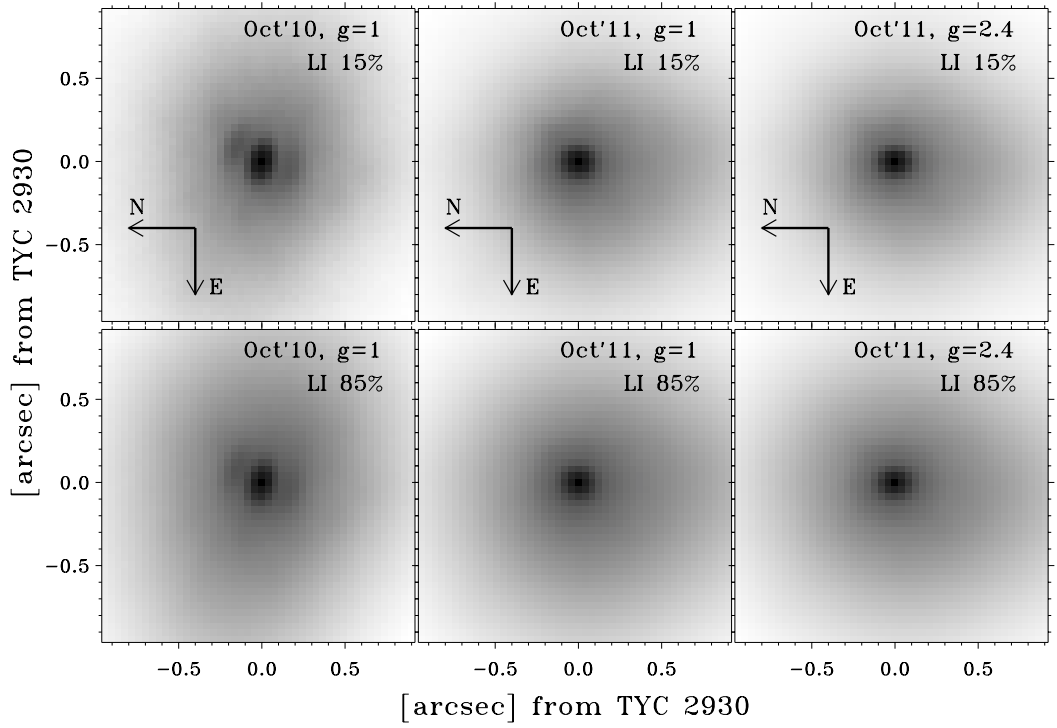}
\caption{Lucky images of TYC 2930 using the best 15 and 85\% of the frames for the 2010 and 2011 observations.  The CCD gain, which changed during the 2011 observations, is labeled by $g$ in the images.  No tertiary companion is detected.\label{luckyimages}}
\end{figure}

The BPs of each frame are then sorted from brightest to faintest, and the best $X$\% are then shifted and added to generate a final image, where $X = \{15, 85\}$.  The effective Strehl ratios for $X = 85$\% are $\{0.036, 0.037, 0.043\}$ for the three image sets, respectively.  Fig.\ \ref{luckyimages} shows composite images for $X = 15 ~ \rm{and} ~ 85$\% of the frames for each set of observations.  The intensities are detector counts on a linear scale after being stacked and normalized by the number of images used in the stacking.  The artifact in the Oct. 2010 frames is a result of imperfect telescope tracking.  No companion is detected at the 3-$\sigma$ level, where $\sigma$ is defined using the procedure in \citet{fem2011} based on the rms of the counts within concentric annuli centered on TYC 2930 and using 8-pixel boxes.  

\begin{figure}
\includegraphics[scale=0.75]{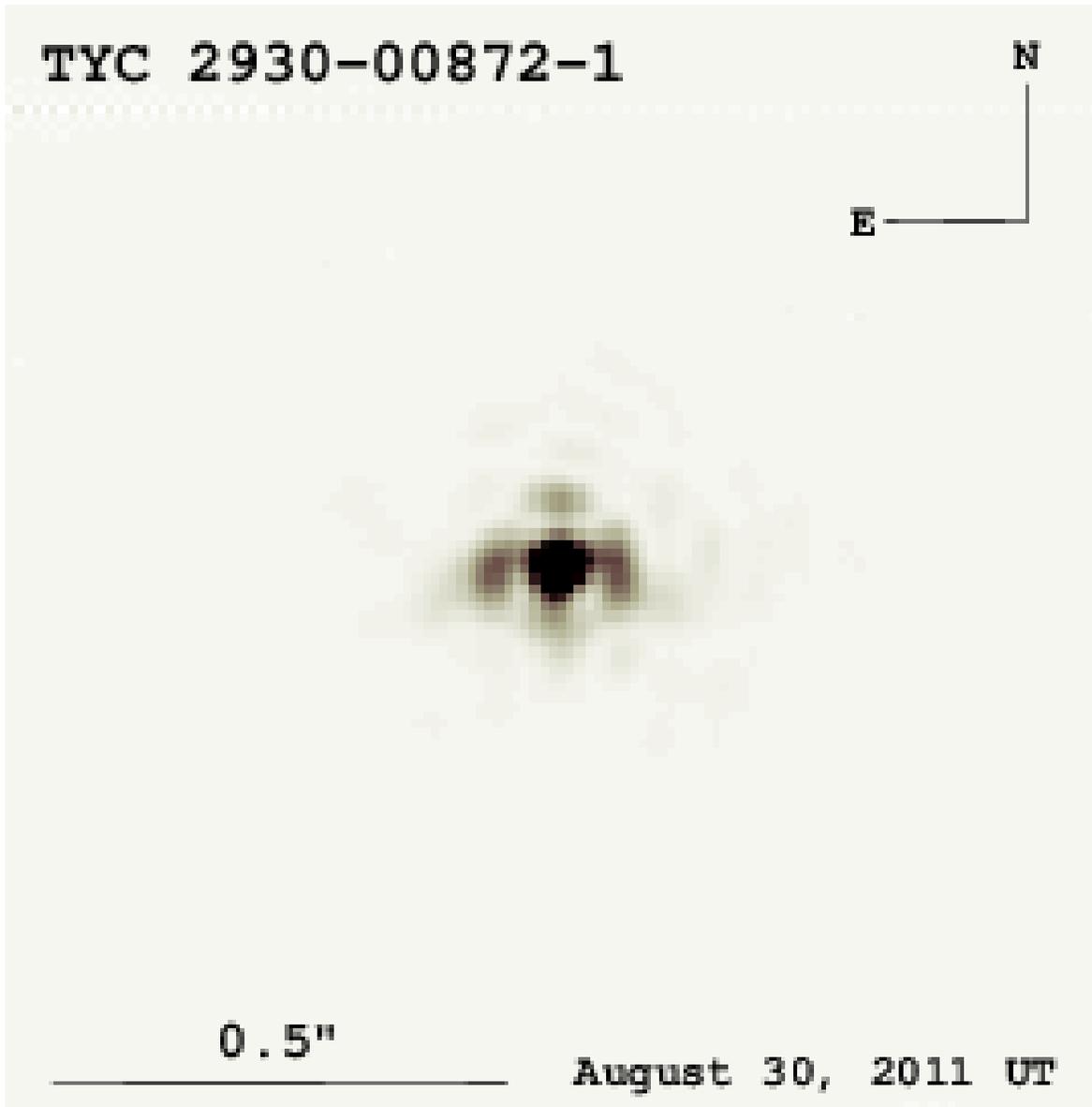}
\caption{Keck AO image of TYC 2930.  No stellar companions are detected at the 3-$\sigma$ level beyond ${\sim}200 ~ \rm{mas} ~ {\simeq}\;30 ~ \rm{AU}$, corresponding to a brightness limit of ${\Delta}m~{\sim}6$.\label{KeckAO}}
\end{figure}

\begin{figure}
\includegraphics[scale=0.58]{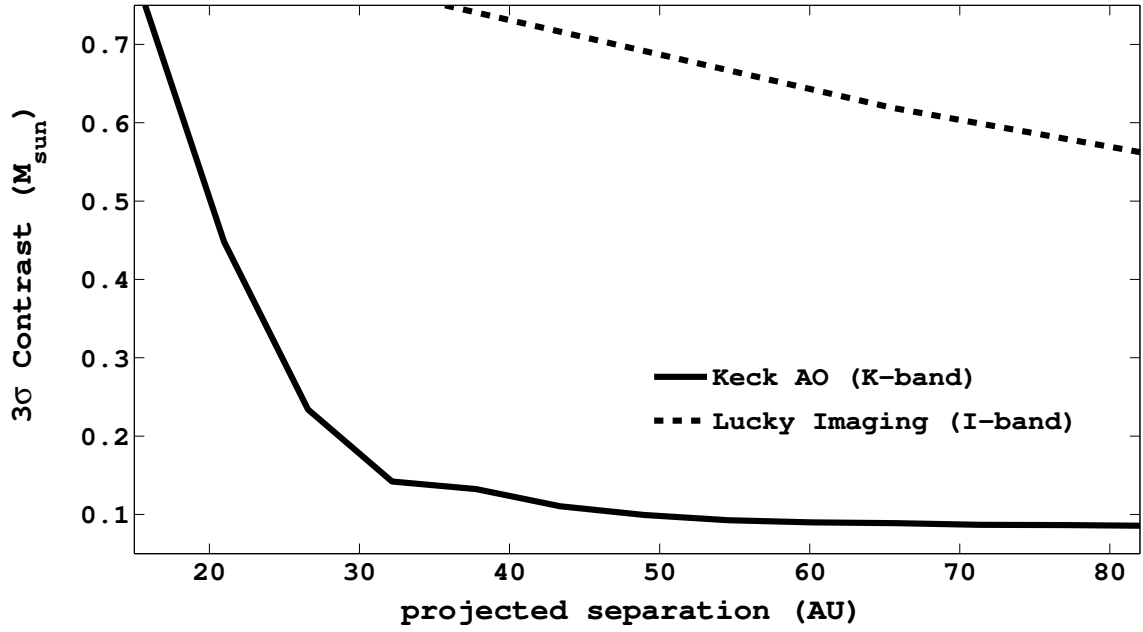}
\caption{Detectability (contrast curve) for the Lucky Imaging and Keck AO images of TYC 2930.  Contrast levels are converted to masses based on \citet{bar2003} models for the Keck band and \citet{gir2002} isochrones for the Lucky Imaging band.  A separation of $50 ~ \rm{AU}$ is $\sim 350 ~ \rm{mas}$ at TYC 2930's Hipparcos-based distance of $\sim 140 ~ \rm{pc}$.\label{KeckAOcc}}
\end{figure}

In addition to the Lucky Imaging, we conducted adaptive optics (AO) imaging to search for any wide stellar companions to TYC 2930.  The Keck AO images were obtained with NIRC2 (PI: K. Matthews) on 2011 August 30 UT.  The observations were conducted in the $K^{\rm{\prime}}$ band using the narrow camera setting, resulting in a plate scale of 9.963 $\rm{mas ~ pix}^{-1}$ \citep{ghe2008}.  The total integration time was 65 sec using a three-point dither pattern.  Images were processed using standard pixel cleaning, flat-fielding and stacking procedures.  Fig.\ \ref{KeckAO} shows the processed Keck AO image; no evidence for wide stellar companions can be seen.  Detectability curves (3-$\sigma$) are calculated as a function of separation from TYC 2930 for both the Lucky Imaging and AO data.  Contrast ratios are converted into mass sensitivities using the \citet{bar2003} models for the Keck band and \citet{gir2002} models for the Lucky Imaging band.  As can be seen in Fig.\ \ref{KeckAOcc}, we can exclude stellar companions at projected separations greater than ${\sim}50~{\rm{AU}}$.  While the Keck constraints are superior compared to the Lucky Imaging constraints, they also rely on a very expensive resource (namely, the Keck telescope).  Since Lucky Imaging can be conducted on much smaller and more readily available telescopes, it is a good resource to use in the search for wide companions in the absence of 10-meter telescope access.

\section{Analysis of the Relative Photometry}
\label{relphot}
\subsection{Summary of Datasets}
The WASP photometric dataset for TYC 2930 consists of 3975 points spanning roughly two years from HJD$'=3831$ to 4571.  The full, detrended WASP dataset has a relatively high weighted rms of $2.9\%$ and exhibits evidence for systematics.  The distribution of residuals from the weighted mean is asymmetric and highly non-Gaussian, showing long tails containing a much larger number of $>3\sigma$ outliers than would be expected for a normally-distributed population.  

We clean the WASP data by adding in quadrature to the photon noise a systematic uncertainty ($\sigma_{\rm{sys}}$) that results in a distribution of residuals closest to the Gaussian expectation.  We reject the largest, error-normalized outlier from the mean flux value and scale the uncertainties by a factor $r$ to force $\chi^2/{\rm dof}=1$, iterating until no more outliers $>4\sigma$ remain.  Although $4\sigma$ is a slightly larger deviation than we would expect based on the final number of points, we adopt this conservative threshold to avoid removing a potential transit signal.  We find $r=0.39$ and $\sigma_{\rm{sys}}=0.0053$, retaining 3731 data points with an rms of $0.71\%$ and $\chi^2/{\rm dof}=1$ (by design).

The Allegheny photometric dataset consists of 1280 points spanning roughly 44 nights from HJD$'$=5596 to 5640.  The weighted rms of the raw light curve is $0.48\%$; this is a factor of $\sim 3$ times smaller than the average fractional photometric uncertainties, indicating that these errors have been overestimated.  Although there is no clear evidence for systematic errors in this dataset, we repeat the identical procedure as with the WASP data for consistency.  We find $r=0.38$ and $\sigma_{\rm{sys}}=0.0011$, with a final rms of $0.42\%$ from 1274 data points.

The KELT dataset consists of 2781 data points spanning roughly 3 years from HJD$'=4107$ to 5213.  The weighted rms of the raw light curve is $0.62\%$, and the mean uncertainty is $0.55\%$, indicating that these are reasonably well estimated.  Nevertheless, for consistency we clean the data in the same way as the other two data sets, finding no outliers $>4\sigma$, $r=1.09$, and $\sigma_{\rm{sys}}=0.0021$, with a final rms of $0.62\%$.

\begin{figure}
\includegraphics[scale=0.8]{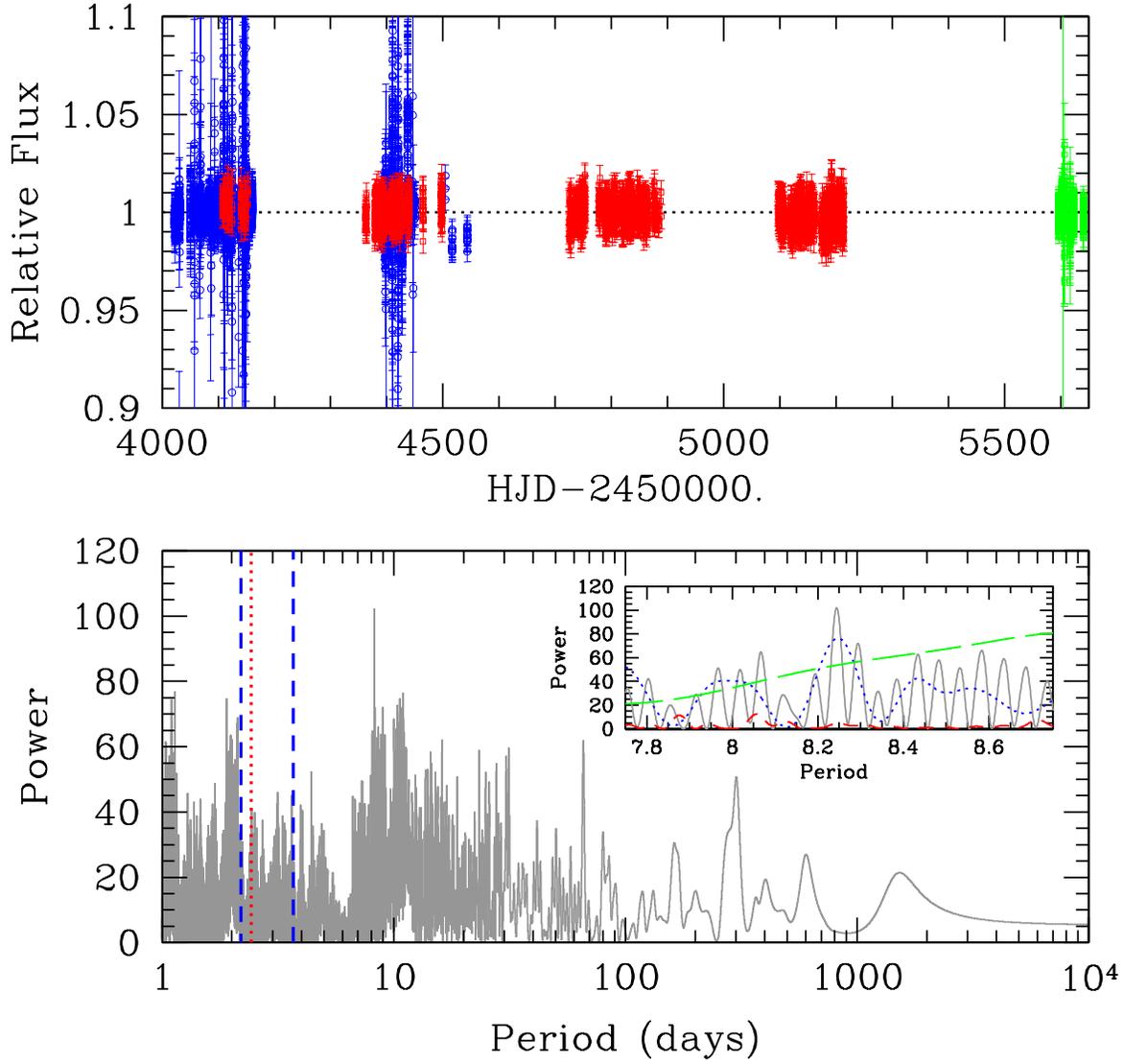}
\caption{Top panel: Cleaned relative photometry of TYC 2930 from WASP (blue), KELT (red), and Allegheny (green).  Bottom panel: Lomb-Scargle periodogram of the combined photometric data.  While a large number of strong peaks are visible, we do not regard these as significant. There is no strong peak at the period of the secondary (vertical red dotted line) or within the estimated $2~\sigma$ period range of the primary's rotational period (vertical blue dotted lines). The inset shows detail of the most significant peak in the combined periodogram (grey), as well as the periodograms for just the WASP (blue dotted), KELT (red short dashed), and Allegheny (green long dashed) datasets.  The peak in the combined dataset arises almost exclusively from the WASP data, and is not confirmed by the KELT data.
\label{powbin}}
\end{figure}

Finally, we combine all the relative photometry after normalizing each individual data set by its mean weighted flux.  The top panel of Fig.\ \ref{powbin} shows the combined data set, which consists of 7786 data points spanning roughly 4.4 years from HJD$'=4022$ to 5640, and has a weighted rms of $0.58\%$.  The resulting light curve is constant to within the uncertainties over the entire time span.  Within the KELT data set, which spans $\sim 3$ years, we find no strong evidence for long-term intrinsic variability at a level $\ga 0.6\%$.

\subsection{Search for Periodic Variability}
We ran a Lomb-Scargle periodogram on the full dataset, testing periods between $1-10^4$ days. The resulting periodogram, shown in the bottom panel of Fig.\ \ref{powbin}, displays a large number of formally significant peaks.  The inferred amplitudes are all $\la 0.1\%$; similar to the level of systematic errors inferred when cleaning the light curves.  A comparison of periodograms performed on the individual data sets demonstrates that the strongest peaks arise from only one dataset, and are not corroborated by the other datasets.  The most significant peak in the combined dataset has a period of $8.24$ days, with a power of $\sim 100$ and an amplitude of $\sim 0.13\%$.  However, as shown in the inset, the signal comes almost entirely from the WASP dataset (blue).  In general, the KELT dataset (red) shows significantly reduced power on all periods $\la 100$ days; the rms of the periodogram in this range is only $\sim$4, as compared $\sim$14 for the combined data set.  Although the different results inferred for different datasets could in principle arise from real variability that is not strictly periodic or persistent, it is more likely that there exist systematics in the data sets in the form of residual correlations on a range of timescales.
 
Restricting attention to periods within $10\sigma$ of that inferred for the companion ($P=2.430420 \pm 0.000006$ days), the maximum power is $\sim$20 with an amplitude of only $\sim$0.06\%, a factor of $\sim$10 times smaller than the rms of the light curve.  Considering an expanded range of periods within $2\sigma$ of the period of the primary as inferred from the $R'_{HK}$ index ($P=2.93 \pm 0.37~{\rm days}$, see \S \ref{geomtides}), the maximum power is $\sim$46 with an amplitude of $\sim$ 0.09\%.  We do not regard these maxima as significant, and conclude that the star does not exhibit periodic variability at either the expected rotation period of the star or the period of the inner companion at a level $\ga 0.1\%$.

\begin{figure}
\includegraphics[scale=0.8]{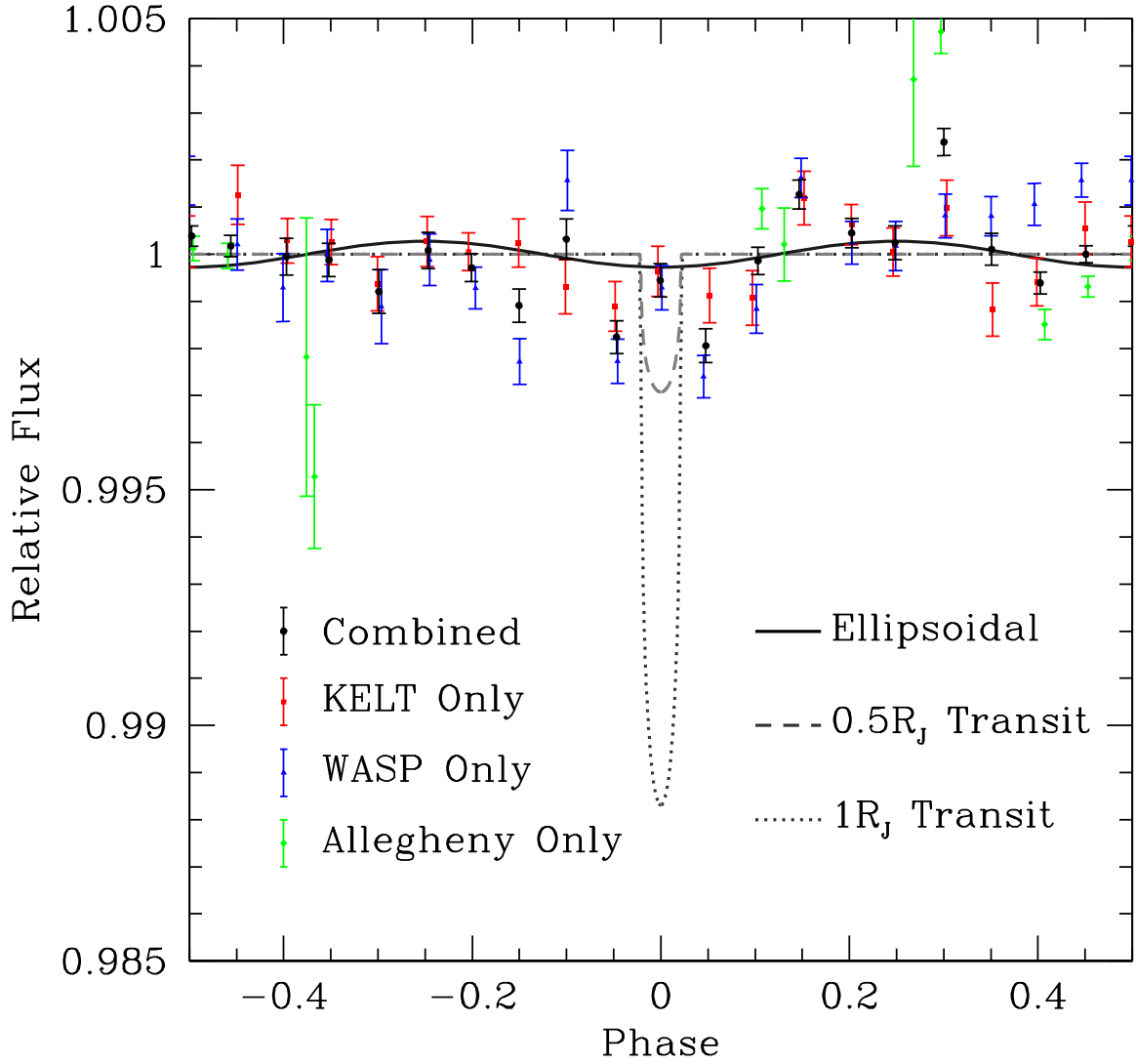}
\caption{Relative photometry folded at the period of secondary, and binned 0.05 in phase.  Phase zero corresponds to the expected time of conjunction (and so of transits for the appropriate inclinations).  Black points are the combined data, blue are WASP, red are KELT, and green are Allegheny.  The grey curves show the expected transit signatures for a companion with radius of $0.5~{\rm{R_{Jup}}}$ (dashed) and $1~{\rm{R_{Jup}}}$ (dotted), assuming an edge-on inclination and the median estimated values of the primary mass and radius.  The solid curve shows the expected signature of ellipsoidal variability assuming an edge-on companion.\label{binall}}
\end{figure}

Fig.\ \ref{binall} shows the combined light curve, folded at the median period and time of conjunction of the companion, ($P=2.430420 \pm 0.000006$ days and $T_C=2454842.2640$), and binned in phase using bins of 0.05. The weighted rms of the binned light curve is $\sim 0.087\%$.  Although the variations are larger than expected from a constant light curve based on the uncertainties $(\chi^2/{\rm dof} \sim$9), we again suggest that these are due to systematic errors in the relative photometry. In particular, the folded, binned KELT light curve (red) shows a somewhat lower rms of $\sim 0.068\%$ with a $\chi^2/{\rm dof} = 1.8$, and is more consistent with a constant flux.  We conclude that there is no strong evidence for variability of TYC 2930 on any time scale we probe.  We can robustly constrain the amplitude of any persistent, periodic variability to be less than $\la 0.1\%$, and we can constrain the amplitude of photometric variability at the period of the companion to $\la 0.07\%$.

Given the estimated stellar mass, the companion period, and the minimum mass, the amplitude of ellipsoidal variability is expected to be:
\begin{equation}
\delta_{\rm ellip} \sim 0.03{\%}\left(\sin{I}\right)\left(\frac{m_2\sin I}{67~{\rm{M_{Jup}}}}\right)\left(\frac{M_1}{1.2~{\rm{M_\odot}}}\right)\left(\frac{P}{2.43~{\rm d}}\right)^{-2}
\end{equation}
with a period of $P/2$ \citep{pfa2008}.  This expected signal is compared to the binned data in Fig.\ \ref{binall}, demonstrating that it is just below the level of detectability.  Since smaller inclinations lead to lower amplitudes, we are unable to constrain the inclination using ellipsoidal viability.

\subsection{Excluding Transits of the Secondary}
\label{transearch}
The probability that a low-mass companion transits can be determined given the orbital parameters from the RV solution \citep{kan2008}.  Assuming a uniform distribution in $\cos{i}$, the a priori transit probability for the secondary is relatively high, $\sim$13\%. However, the light curve folded on the ephemeris of the inner companion shows no evidence for a transit at the expected time of conjunction, with an upper limit to the depth of any putative transit of $\la 0.2\%$. In contrast, the central transit of a Jupiter-sized companion would be expected to have a depth of $\delta \sim (r/R_1)^2 \sim 0.9\%$ and a duration of $\sim$0.042 in phase.  We conclude that our observations rule out such a transiting companion with high confidence.

We perform a quantitative search for transit signals using a method similar to that described in \citet{fle2010}.  We use the distributions of the secondary period $P$, semiamplitude $K$, and time of conjunction $T_c$ from the MCMC RV analysis, setting the eccentricity of the secondary to zero for simplicity.  For each link in the MCMC chain, we draw a value of $T_{\rm eff}$, $\log g$, and $\rm{[Fe/H]}$ for the primary from Gaussian distributions, with means and dispersions given in Table \ref{stellarprops}, and determine the primary's mass $M_1$ and radius $R_1$ using the \citet{tor2010} relation on these values.

We then draw a value of $\cos i$ from a uniform distribution\footnote{Formally, this assumes a prior on the companion mass $m_p$ that is uniform in $\log{m_p}$.}, and use the resulting values of $P$, $K$, $M_1$, $R_1$ and $i$ to determine the secondary mass $m_p$, semimajor axis $a$, and impact parameter of the secondary orbit $b \equiv a\cos{i} ~ R_1^{-1}$.  Finally, adopting a radius for the companion, $r$, we determine if the companion transits, and if so we determine the properties of the light curve using the routines of \citet{man2002}.  We assume quadratic limb darkening and adopt coefficients appropriate for the $R$ band from \citet{cla2003}, assuming solar metallicity and the values of $T_{\rm eff}$ and $\log g$ listed in Table \ref{stellarprops}.  For reference, Fig.\ \ref{binall} shows the predicted transit signatures for the median values of the physical parameters and $r=0.5~{\rm{R_{Jup}}}$ and $1~{\rm{R_{Jup}}}$.  We fit the predicted transit light curve to the combined photometric data, and then compute the $\Delta \chi^2$ between the constant flux fit and the predicted transit model.

Our best-fit has a $\Delta \chi^2=-19.8$, which we do not consider significant.  We find similar or larger improvements in $\chi^2$ when we consider arbitrary phases for the transit and when we consider ``anti-transits" \citep[signals with the same shape as transits but corresponding to positive deviations, see][]{bur2006}.  As before, these formally significant signals likely arise from systematics in the photometric data.  We conclude there is no evidence for a transit signal in the combined data.

\begin{figure}
\includegraphics[scale=0.7]{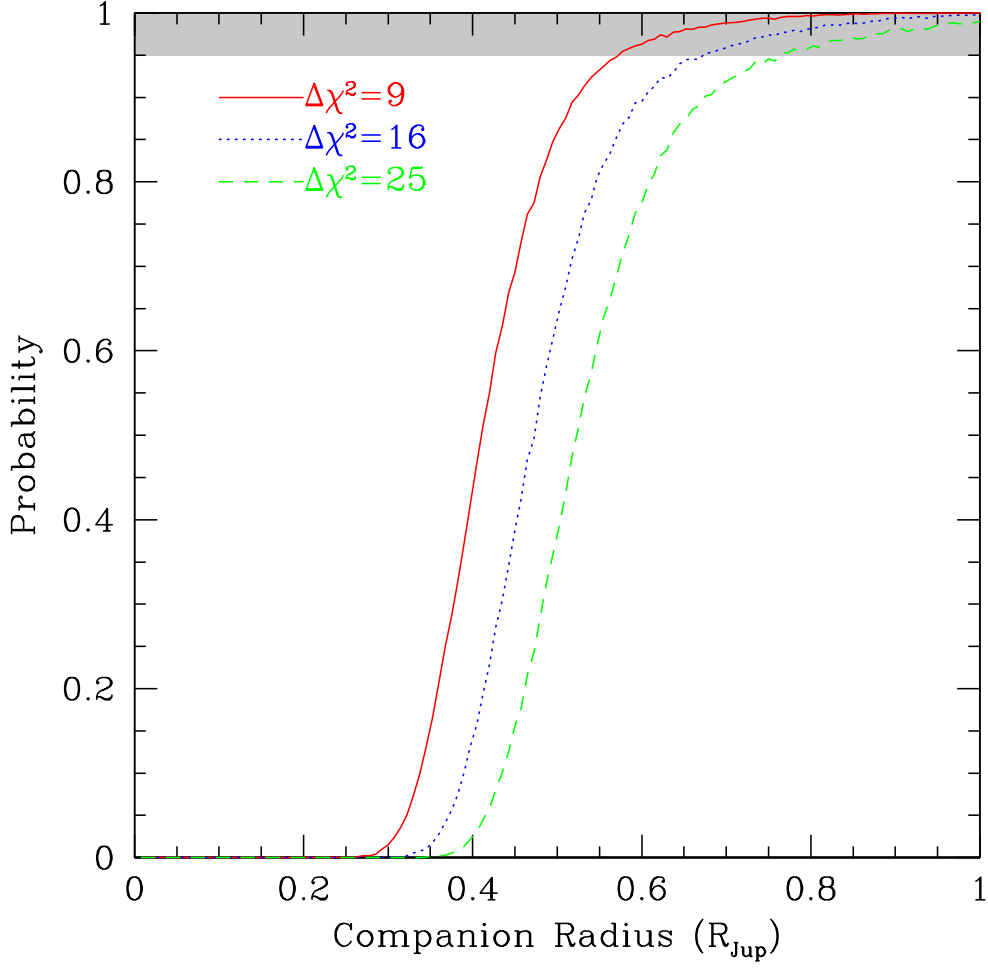}
\caption{Probability that transits of a companion are excluded at levels of $\Delta\chi^2=9, 16, 25$ based on the analysis of the combined WASP, KELT, and Allegheny photometric data sets, as a function of the radius of the companion.  Transits of companions with radius $r\ge 0.7~{\rm{R_{Jup}}}$ can be excluded at the 95\% confidence level.\label{exctrans}}
\end{figure}

Given that we do not detect a transit signature, we can also use this procedure to determine the confidence with which we can rule out transits of a companion with a given radius.  This is just given by the fraction of the steps in the Markov Chain for which the companion transits and produces a transit signature with a $\Delta\chi^2$ relative to the fixed constant flux greater than some threshold.  We consider thresholds of $\Delta\chi^2 = 9$, 16, and 25.  The resulting cumulative probability distributions for a range of companion radii are shown in Fig.\ \ref{exctrans}.  Given the systematics in the data, we consider thresholds of $\Delta\chi^2 \ga 25$ to be robust, and thus conclude that transiting companions with $r \ga 0.75~{\rm{R_{Jup}}}$ are likely ruled out at the $\sim 95\%$ confidence level.  The models of \citet{bar2003} predict radii of $\ga 0.8~{\rm{R_{Jup}}}$ for brown dwarfs of $m_p\sim 60~{\rm{M_{Jup}}}$ and ages of $\la 5~{\rm Gyr}$. Given the upper limit of the age of the primary of $\sim 2~{\rm Gyr}$, we can therefore essentially rule out non-grazing transits of the companion.

\section{Constraints on System Geometry and Tidal Analysis}
\label{geomtides}
The rotational period of a star can be estimated from an empirical relationship between the Rossby number and $\log{R^{\prime}_{HK}}$ \citep{noy1984}.  The Rossby number $R_{0} = P ~ \tau_{c}^{-1}$, where $P$ is the rotation period of the star and $\tau_{c}$ is the convective turnover time.  In this work, we use the relationship as quantified by \citet{mam2008}.  We estimate the convective turnover time based on the relationship between $R_{0}$ and $\left(B-V\right)$ from \citet{noy1984}.  We find an expected rotational period based on the measured $\log{R^{\prime}_{HK}}$ of $P = 2.93 \pm 0.37$ days.  The uncertainty in the period includes the uncertainty in the \citet{mam2008} relationship, as well as an adopted uncertainty in $R^{\prime}_{HK}$ of $0.2 \times 10^{-5}$ based on the observed variability of the most active stars in the \citet{lov2011} sample.  The latter uncertainty accounts for the fact that the \citet{mam2008} relation applies to a time-averaged $R^{\prime}_{HK}$, while we have a single epoch measurement.  The modest upper limit on the rotation rate of $v\sin I < 6 ~ \rm{km ~ s^{-1}}$, the fact that the estimated rotation rate from the \citet{mam2008} relation is close to the orbital period of the inner companion, and the expected equatorial rotation velocity of $22.68 \pm 3.12 ~ \rm{km ~ s^{-1}}$ if the primary was rotating at the inner companion's orbital period and had a stellar spin axis oriented edge-on, all suggest that the primary's spin axis is inclined relative to our line-of-sight, in which case the inclination is constrained to be $I = 15.0_{-6.2}^{+7.3}$ degrees.

We can use the reasonably strong upper limit on TYC 2930's photometric variability to place additional limits on the inclination of the star and its companion. The lack of photometric variability is somewhat surprising, given the spectroscopic indications that the star is relatively active.  An estimate of the expected photometric variability can be obtained using the relationship in \citet{har2009} between $R_{0}$ and photometric amplitude.  Based on the estimated $R_{0}$ from the \citet{mam2008} $R_{0}$-$\log{R^{\prime}_{HK}}$ relationship, we would expect a photometric amplitude of $\sim$0.8\%.  Figure 17 in \citet{har2009} indicates that there exists substantial scatter about this relation, which is likely partially suppressed for values of $R_{0} \ga 0.4$ due to incompleteness. Extrapolating the observed scatter at lower values of $R_{0}$ (which are less affected by incompleteness), we expect a photometric amplitude in the range of $\sim$0.2-2\%.  Thus, the fact that the observed amplitude is a least a factor of $\sim$3 times lower ($\la 0.07\%$) suggests that this star is either surprisingly photometrically quiet given its spectroscopic activity indicators, or is it being viewed nearly pole-on.

The expectation of a synchronous rotation rate for the primary is reinforced by consideration of the tidal evolution of the primary and secondary. To analyze this effect, we employed the ``constant-phase-lag'' (CPL) tidal model in which the location of the tidal bulge on the two bodies lies at a constant angle from the line connecting the centers of the two bodies. Frictional forces inside the two objects prevent perfect alignment, which leads to energy dissipation and transfer of angular momentum. Hence the system will evolve with time \citep[see e.g.][]{gol1966,fer2008}. Two outcomes are possible: arrival at the ``double synchronous'' state, in which the obliquities are normal to the orbital plane, and the two spin frequencies equal the orbital frequency, or the two bodies merge \citep{cou1973}. Here we use the CPL model presented in \citet{fer2008}, with the numerical methods described in \citet[][App.~D]{bar2012}.

In the CPL model, the tidal effects scale with the ``tidal quality factor'' $Q_*$. This parameter is poorly constrained in stars and brown dwarfs with values ranging from $10^5$--$10^9$ \citep[e.g.][]{lin1996,mat2008}; therefore we explore the timescale to reach the double synchronous state in this range. The masses and radii of the two objects are also important, and although we have constraints on the two masses and the primary's radius, the secondary's radius is unknown. We adopt a radius of 1 Jupiter mass, in line with theoretical models \citep{bar2003}.

For compact binaries, tidal evolution during the pre-main sequence phase is also important \citep{zah1989,kha2011,gom2012}, since the radii are larger. As tidal effects scale with radius to the fifth power, the results can be dramatic. Most stellar binaries with periods less than 8 days are on circular orbits and probably near the double synchronous state \citet{zah1989}, suggesting the TYC 2930 system could also have reached this state early in its history. On the other hand, the inner companion could have arrived at its orbit after the radial contraction phase, perhaps via a gravitational scattering event \citep{heg1975}, or due to Kozai-Lidov interaction followed by tidal circularization \citep{fab2007}. Therefore, in order to be as conservative as possible, we will ignore any evolution during the pre-Main Sequence. We therefore consider our timescales to reach double synchronization to be upper limits.

\begin{figure}
\includegraphics[scale=0.65]{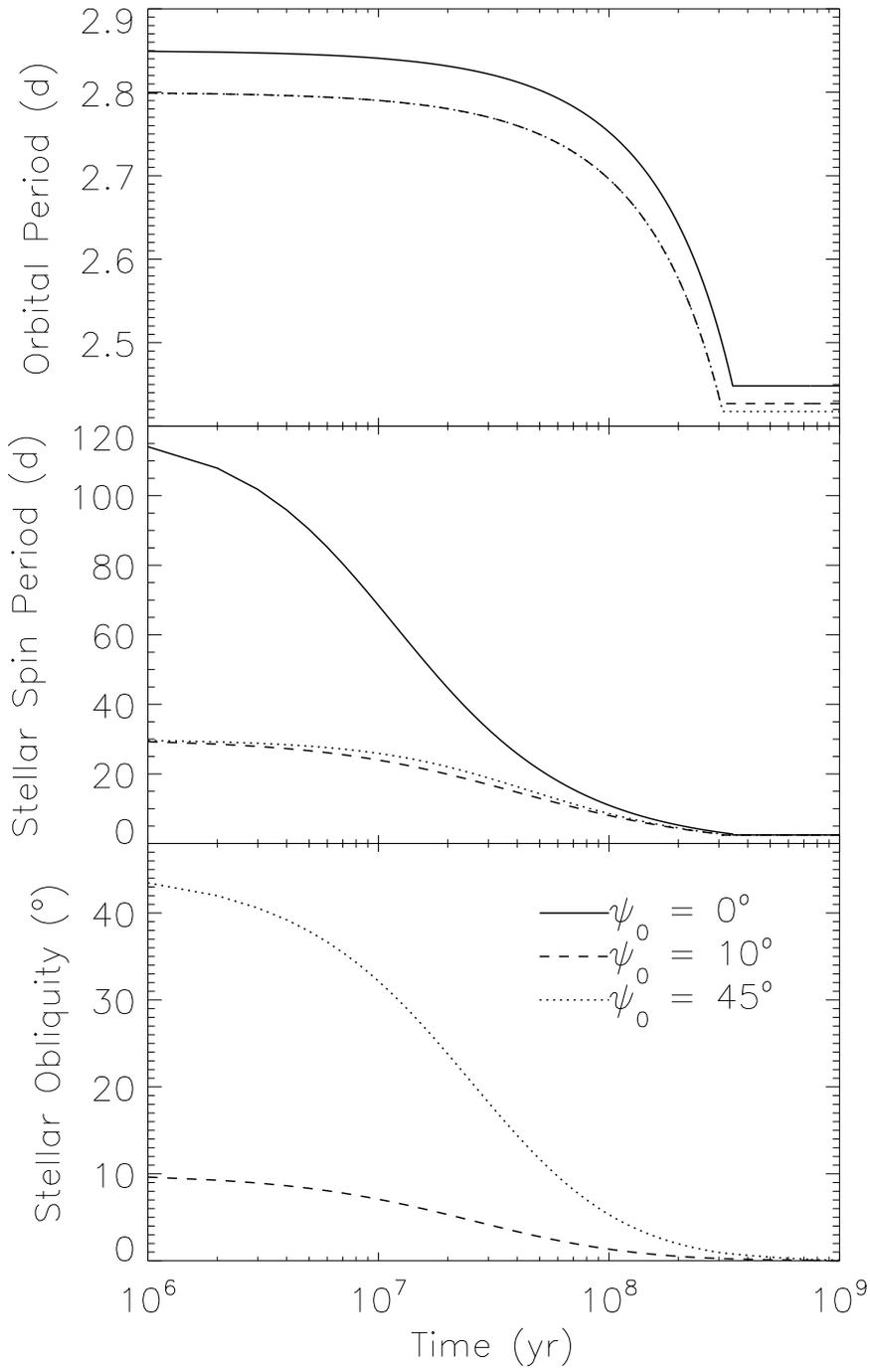}
\caption{Tidal evolution of the TYC 2930 system. Curves represent assumed initial stellar obliquity values of $\psi_0 = 0^\circ$ (solid), $10^\circ$ (dashed), and $45^\circ$ (dotted). {\it Top:} Evolution of the orbital period. {\it Middle:} Evolution of the stellar spin period. {\it Bottom:} Evolution of the stellar obliquity.
\label{tidalevolfig}}
\end{figure}

In Fig.\ \ref{tidalevolfig} the evolutions of the orbital period, stellar spin period and stellar obliquity, $\psi$, are shown for three different cases of initial obliquities: $\psi_0 = 0^\circ$ (solid curves), $\psi_0 = 10^\circ$ (dashed curves) and $\psi_0 = 45^\circ$ (dotted curves). For each case, the initial spin period is 30 days, the secondary has a mass of $100 ~ \rm{M_{Jup}}$, the initial orbital period is 2.8 days, and the stellar tidal quality factor $Q_*$ is $10^7$.  In each model, the orbital and spin periods reach $\sim 2.43$ days and become locked after $\sim 300$~Myr. The obliquity can evolve slightly longer, but with negligible effect on the spins and orbit. Double synchronization requires $\psi = 0^\circ$ and hence the timescale to reach that state is larger when the initial $\psi$ is nonzero.

The initial spin period of the star is unknown, and therefore we tested values larger than 30 days. For the extreme case of a 120 days period and $\psi_0 = 45^\circ$, the time to reach double synchronization is about 3 times longer than the cases shown in Fig.\ \ref{tidalevolfig}. For this initial configuration, an initial orbital period of 2.85 days produces a better match to the observed system.

\begin{figure}
\includegraphics{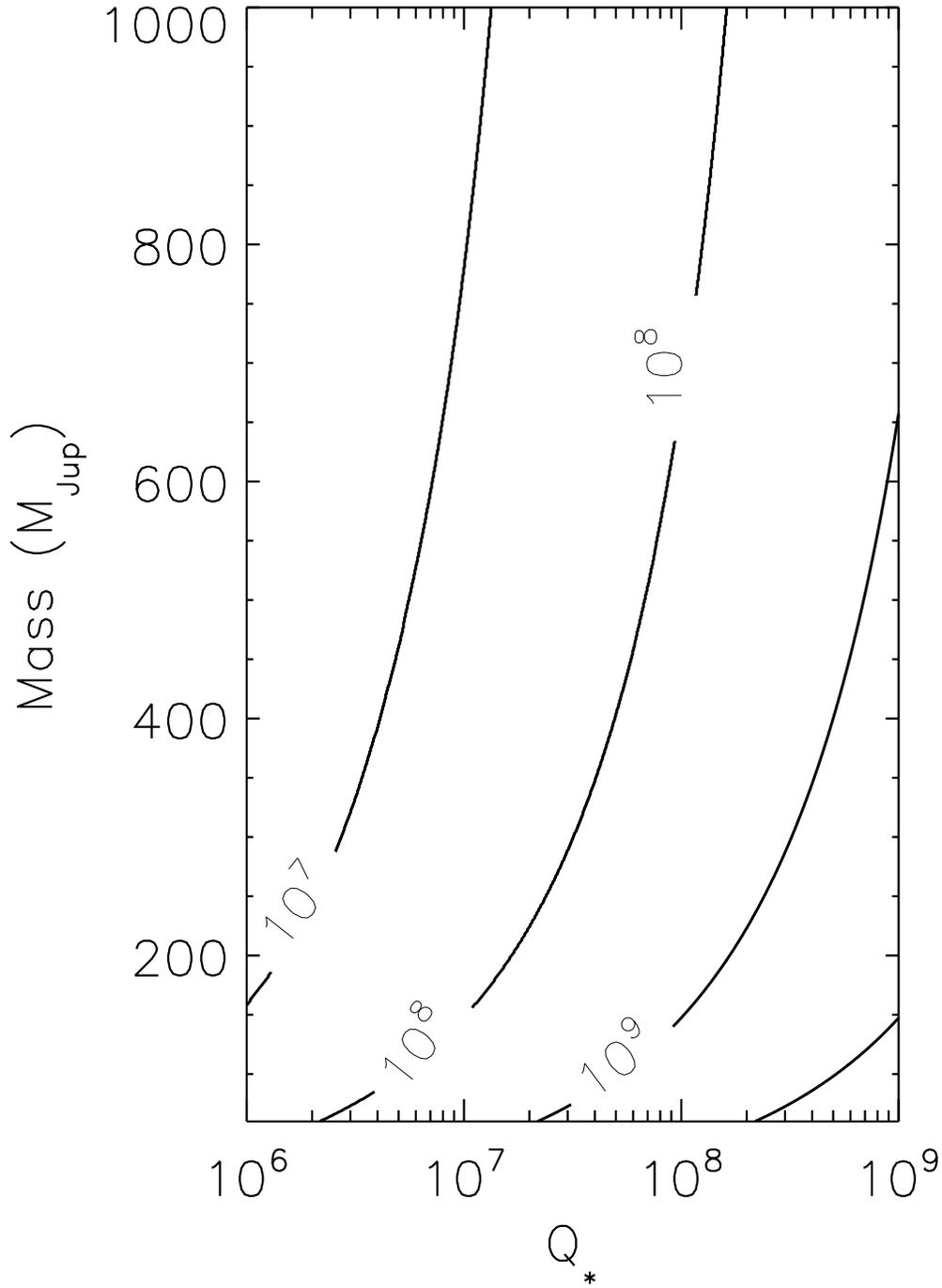}
\caption{Time (years) for TYC 2930 to reach double synchronization. The star initially has no obliquity and a rotation period of 30 days. The companion is tidally locked. No non-tidal effects, such as an early epoch of radial contraction, are included.
\label{dblsyncfig}}
\end{figure}

We expanded our analysis to a range of $M_2$ and $Q_*$ to determine which values predict the double synchronized state on a timescale less than or equal to the system age of $< 2$ Gyr. Initially, we set the primary spin period to 30 days and $\psi_0$ to $0^\circ$ (as shown in Fig.\ \ref{tidalevolfig}, different choices do not affect our results), the secondary's spin period was always tide-locked, and the initial orbit was circular with a period of 2.8 days. We then integrated forward a suite of configurations in the ranges $60 \le M_2 \le 1000$~$\rm{M_{Jup}}$ and $10^6 \le Q_* \le 10^9$ and calculated the timescale to reach double synchronization. We considered this state achieved when the stellar spin period was within 10\% of the orbital period. Fig.\ \ref{dblsyncfig} shows the timescale for double synchronization as a function of inner companion mass and $Q_{*}$.  For the minimum mass, double synchronization will occur within the nominal 2~Gyr lifetime if $Q_* \lesssim 3 \times 10^7$, while at $M_2 = 300$~$\rm{M_{Jup}}$, $Q_*$ must be $<4 \times 10^8$. Our tidal analysis strongly favors the double synchronous state.

\section{A Posteriori Distributions of the True Mass}
\label{massdistsection}
The a posteriori distribution of the true mass of the companions given our measurements depends on our prior distribution for the mass of the companion, or, roughly equivalently, our prior on the mass ratio.  As a rough illustration, if we assume a prior that is uniform in the logarithm of the true mass of the companion, then the distribution of $\cos i$ will be uniform.  Therefore, the median $\cos i \simeq 0.5$, and thus the median $\sin i \simeq 0.866$.  Thus we have,
\begin{equation}
0.75^{3/2} M_j^3  - {\cal M}_j (M_1+M_j)^2 = 0,
\label{eqn:mfsol}
\end{equation}
where $j=2,3$ refers to the secondary and tertiary, respectively.  For the secondary, ${\cal M}_2=1.6711 \times 10^{-4}$ and $M_1=1.21~\rm{M_\odot}$; the solution is $M_2 \simeq 0.0752~\rm{M_\odot}$ or $\sim 78.75~\rm{M_{Jup}}$.  This result is roughly $M_{2,\rm min}/\sqrt{0.75}$, but not exactly.  For the tertiary, the median mass is $M_3 \sim 0.48~\rm{M_\odot}$.

More generally, for other priors, $\cos i$ is not uniformly distributed.  We adopt priors of the form:
\begin{equation}
\frac{dN}{dq} \propto q^{\alpha}
\label{eqn:qprior}
\end{equation}
where $q$ is the mass ratio between the companion and the primary, and $\alpha=-1$ for the uniform logarithmic prior discussed above.  To include this prior, we draw a value of $\cos i$ from a uniform distribution for each link the MCMC chain, but then weight the resulting values of the derived parameters for that link (i.e., the companion mass $m$) by $q^{\alpha+1}$.  In addition, based on the analysis of the relative photometry in \S \ref{transearch}, we exclude transiting configurations of the secondary such that $b \equiv a\cos{i_2} ~ R_1^{-1}<1$.

For $q>0$, the a posteriori distribution does not converge, i.e.,
there is finite probability at infinitely large true masses.  However,
we can rule out nearly equal mass ratio main-sequence companions by the lack of any
evidence of additional light beyond that from the primary, in particular from the lack of a
second set of spectral lines and from the shape of the spectral energy
distribution.  As we describe below,
both of these constraints place an upper limit on the mass of any
main-sequence companion to $\la 1 ~ \rm{M_\odot}$.  We adopt this limit by
giving zero weight to inclinations of the secondary and tertiary such that $M_{2,3}\ge 1 ~ \rm{M_\odot}$.  In
doing so, we implicitly assume that neither of these companions are stellar remnants.

The optical/near-infrared colors of the
primary are well-fit by a model for the spectral energy distribution
appropriate to a star with the effective temperature and surface
gravity we measure from the spectrum (see Fig.\ \ref{sedfit}).  A
sufficiently massive main-sequence companion would contribute near-IR
flux in excess of that predicted from the primary.  We can therefore
use the lack of excess flux to constrain the mass of the secondary
(and tertiary).  We generate a set of two-component SEDs using \citet{bar1998,bar2003} models, adopting stellar parameters for the primary from Table \ref{stellarprops} and models corresponding to tertiary masses from 0.2 to 1.2 $\rm{M_{\odot}}$.  We then fit the observed fluxes with these two-component SEDs, allowing distance and line-of-sight extinction to be free-parameters.  For each tertiary mass, we calculate a $\chi^2 = \chi^2_{\rm{SED}} + \chi^2_{\rm{dist}}$, where $\chi^2_{\rm{SED}}$ is based on the observed fluxes and $\chi^2_{\rm{dist}}$ is based on the Hipparcos parallax.  We compute a ${\Delta}\,{\chi^2} = \chi^2_{\rm{2comp}} - \chi^2_{\rm{1comp}}$, where $\chi^2_{\rm{1comp}}$ is the $\chi^2$ for the single-component (i.e., primary-only) SED fit.  We then use these ${\Delta}\,{\chi^2}$ to determine an upper limit of the tertiary mass by rejecting those tertiary masses that yield ${\Delta}\,{\chi^2} > 0$, finding $M_3 \lesssim 1 ~ \rm{M_{\odot}}$.  A similar mass constraint is derived by visual inspection of the cross-correlation functions from the spectroscopic observations.  Assuming none of the components are evolved, mass ratios of $\sim$ 0.8 can be excluded at the maximum RV separation between the primary and tertiary $\left(\sim 5 ~ \rm{km ~ s^{-1}}\right)$ due to a lack of asymmetry in the correlation peaks, once again constraining $M_3 \lesssim 1 ~ \rm{M_{\odot}}$.

Finally, as discussed in \S \ref{geomtides}, the rotational period of the primary $P_*$ as estimated from the $R'_{HK}$ index is within $1.5\sigma$ of the period of the secondary, suggesting that the primary spin may be synchronized to the companion orbit.  This hypothesis is corroborated by our tidal analysis, which suggests that for reasonable values of the stellar $Q_*$ this system should reach a double-synchronized state in which the obliquities are aligned with the orbit within 100~Myr, which is considerably less than our estimate of the age of the system.  Therefore, adopting this synchronized/coplanar assumption, we have $I=i_2$ and $P_*=P_2$. We can use these assumptions, combined with the constraints on $v\sin I = 3.8_{-2.8}^{+1.9}~{\rm km~s^{-1}}$ from the spectroscopic analysis of the primary, to constrain $i_2$ and thus the true mass of the companion.  For each link the MCMC chain, we use the value of $i_2$, $R_1$, and secondary period $P_2$ to estimate $v\sin I = v\sin i_2 = (2\pi R_1/P_2)\sin i_2$.  We then multiply the weight of that chain by the additional factor $\exp[-0.5(v\sin I-3.8~{\rm km~s^{-1}})^2/\sigma_{v\sin I}^2]$ where $\sigma_{v\sin I}=2.8~{\rm km~s^{-1}}$ for $v\sin I<3.8~{\rm km~s^{-1}}$ and $\sigma_{v\sin I}=1.9~{\rm km~s^{-1}}$ for $v\sin I>3.8~{\rm km~s^{-1}}$.

\begin{figure}
\includegraphics[scale=0.7]{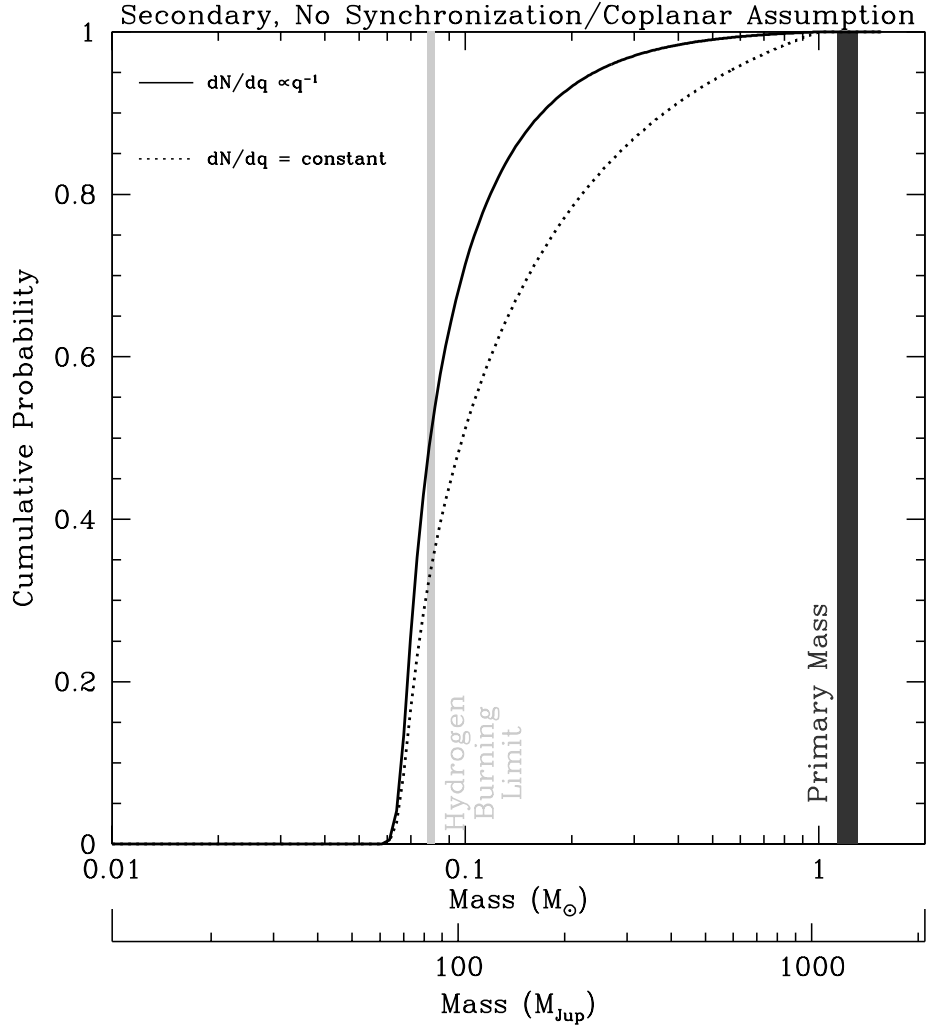}
\caption{The cumulative a posteriori probability of the true mass of
the secondary, for two different priors on the
mass ratio $q$: $dN/dq \propto q^{-1}$ (uniform in $\log{q}$), and
$dN/dq$ = constant (uniform in $q$).  Companion
masses $\ge 1 ~ \rm{M_\odot}$ and transiting configurations are excluded.\label{mass10}}
\end{figure}

\begin{figure}
\includegraphics[scale=0.7]{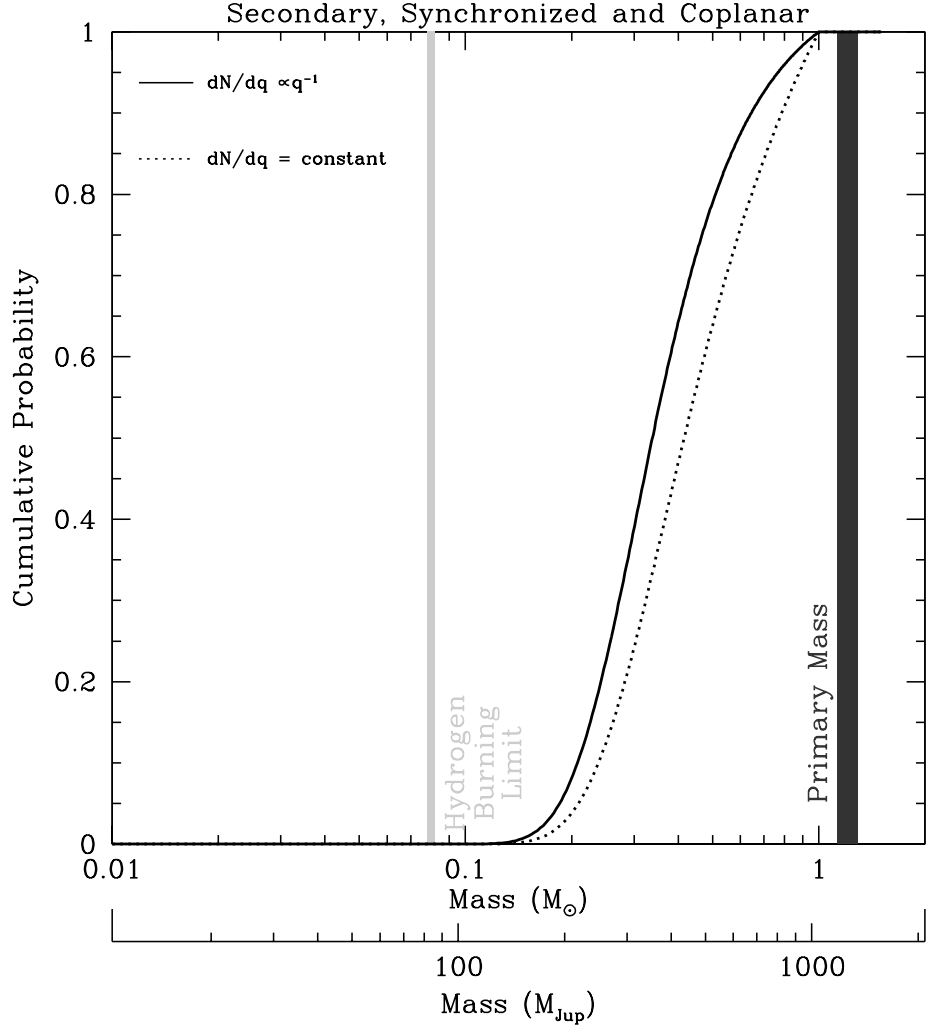}
\caption{The cumulative a posteriori probability of the true mass
of the secondary under the same set of assumptions as Figure
\ref{mass10}, but assuming that the primary and secondary are
synchronized and the stellar obliquity has been damped to zero, and
thus the inclination of the star $I$ as determined from the observed
velocity broadening $v_*\sin I$ and stellar rotation period $P_*$ is equal to
the inclination of the orbit of the secondary $i$.\label{mass11}}
\end{figure}

\begin{figure}
\includegraphics[scale=0.7]{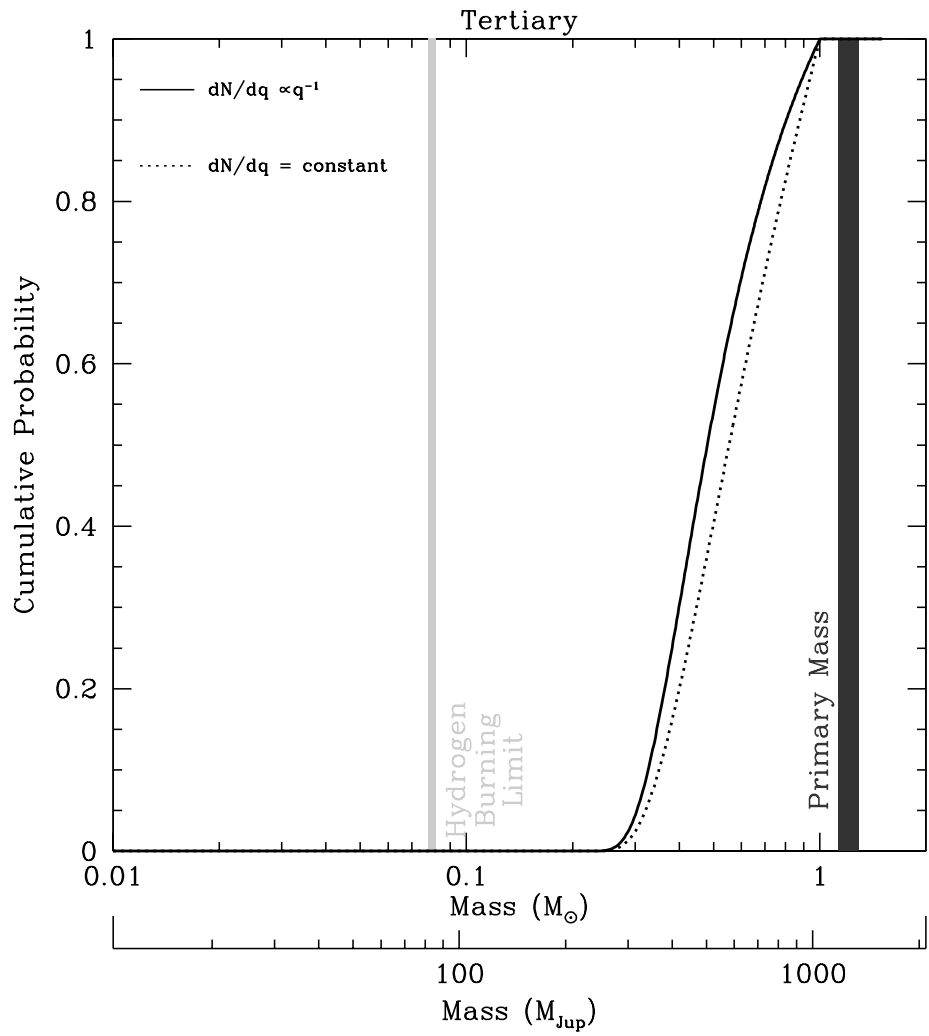}
\caption{The cumulative a posteriori probability of the true mass
of the tertiary under the same set of assumptions as Figure
\ref{mass10}, except that transiting configurations are not
excluded.\label{mass20}}
\end{figure}

Figs.\ \ref{mass10}, \ref{mass11}, and \ref{mass20} show the
resulting cumulative a posteriori distributions of the true masses of
the companions under several sets of assumptions.  These figures
illustrate the effect of adopting various priors and constraints on
our inferences about the nature of these companions.  Tables
\ref{apriorimass1} and \ref{apriorimass2} list the median and 68\%
confidence intervals on the true masses of the companions.  We will
discuss each model in turn.

Fig.\ \ref{mass10} shows the results for the secondary, for
$\alpha=-1$ (uniform logarithmic prior on $q$) and $\alpha=0$ (uniform
linear prior on $q$), assuming the constraint from the lack of
transits and the upper limit on the mass of $\rm{M_\odot}$
from the lack and evidence of light from the companion.  The 
constraint that $M_{2} \le 1 ~ \rm{M_\odot}$ makes little difference for $\alpha=-1$, because the
companion is unlikely to be sufficiently massive to contribute a
significant amount of flux.  For $\alpha=0$, the prior on the
companion mass is more heavily weighted to larger masses, and as a
result this constraint does affect the high-mass tail of the
probability distribution; nevertheless, the inferred median masses in
the two cases are similar.  Under these assumptions, we would conclude that the
secondary is most likely a low-mass stellar or brown dwarf companion seen at
a moderate inclination of $\sim 40-60^\circ$, with a median mass just
above the hydrogen-burning limit $M_2 \sim 0.08-0.1~\rm{M_\odot}$.

Assuming synchronization and coplanarity changes the conclusion about
the nature of the secondary dramatically, given that the
spectroscopically-measured $v\sin I$ suggests a relatively low
inclination for the primary of $\sim 15^\circ$.  Fig.\ \ref{mass11}
shows the inferred cumulative distribution for the secondary, for
$\alpha=0$ and $\alpha=-1$ and assuming the upper limit on the mass.  
The results for the two priors are broadly similar: under the
synchronization assumption the secondary is most likely a mid M dwarf with
$M_2\sim 0.3-0.4~\rm{M_\odot}$ with a nearly pole-on orbit with $i_2 \sim
10-13^\circ$.

Fig.\ \ref{mass20} shows the cumulative distributions of the true
mass of the tertiary.  The mass of the tertiary is limited from below
to be $\ga 0.2~\rm{M_\odot}$ by the measured mass function, and from above
to be $\la 1 ~ \rm{M_\odot}$ by the lack of light from the
companion.  The choice of prior on $q$ has a relatively weak affect on
the distribution of masses within this range.  The tertiary is most likely
to be an early M dwarf with a mass of $\sim 0.5-0.6~\rm{M_\odot}$ with a
moderate inclination of $i_3 \sim 60-70^\circ$.  

\begin{figure}
\includegraphics[scale=0.7]{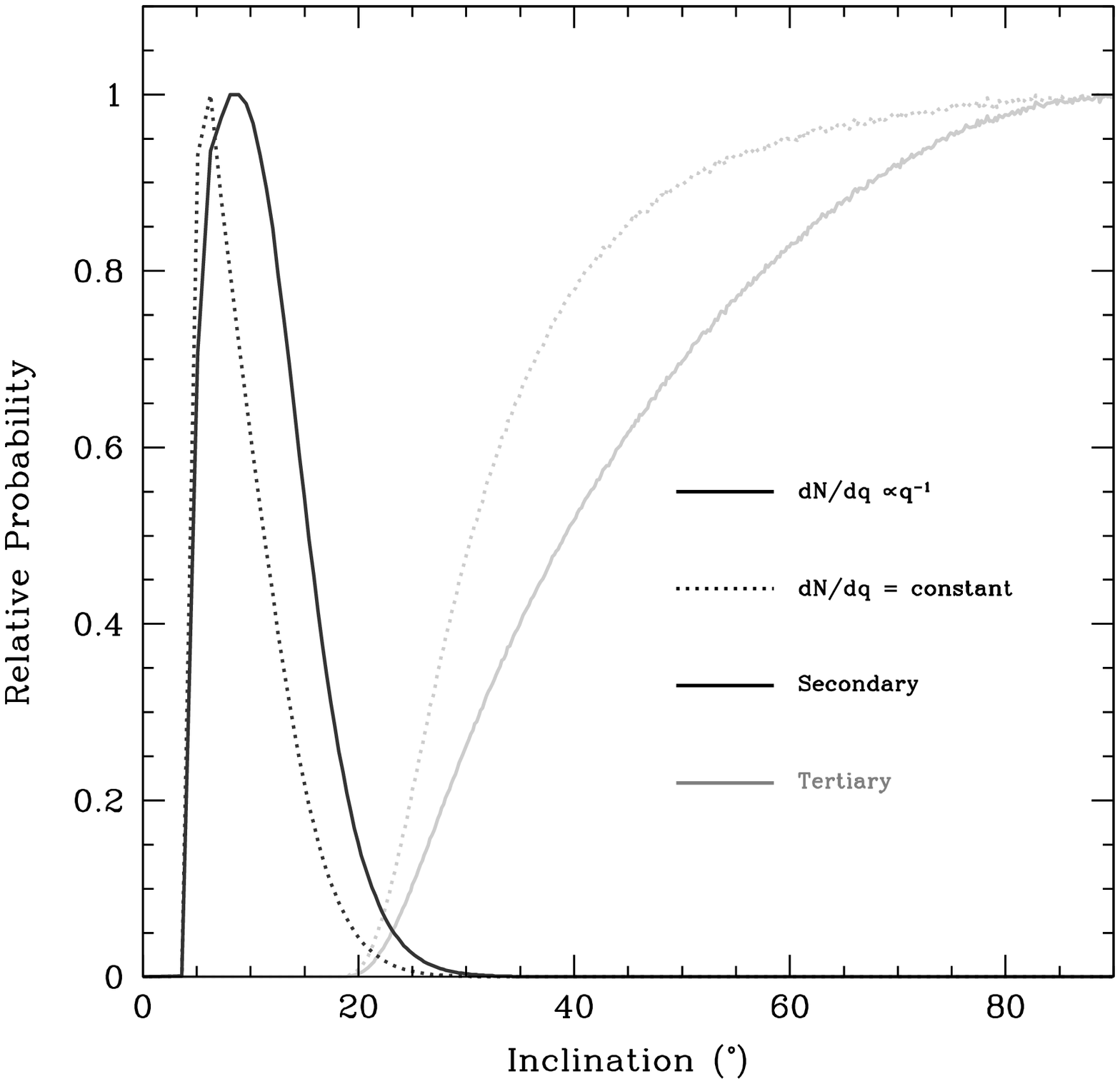}
\caption{The a posteriori probability densities for the
inclinations of the tertiary and secondary.  These curves have been
arbitrarily normalized by their peak probability density.  The black
lines show the probabilities for the secondary assuming the tidal
synchronization/coplanarity, for two different priors on the mass
ratio $q$, $dN/dq \propto q^{-1}$ (uniform in $\log{q}$, solid), and
$dN/dq$ = constant (uniform in $q$, dotted).  The grey lines show
the probabilities for the tertiary, for the same priors on the mass
ratio.  In all cases, we exclude companion masses $\ge 1 ~ \rm{M_\odot}$.\label{ib}}
\end{figure}

Adopting the synchronization/coplanarity assumption, we infer that the
secondary is seen nearly pole-on, whereas from the lack of evidence for additional
light from a companion, we infer that the tertiary cannot be very massive and thus
cannot have a low inclination. Fig.\ \ref{ib} shows a posteriori
probability densities for the inclinations of the tertiary and
secondary under these assumptions.  The $95\%$ confidence level (c.l.)
upper limit on the secondary inclination including the
synchronization/coplanarity constraint and flux constraint is
$23.6^\circ$ ($\alpha=-1$) and $21.2^\circ$ ($\alpha=0$). On the other
hand, the $95\%$ c.l. lower limit on tertiary inclination including
the flux constraint is $30.1^\circ$ for $\alpha=-1$ and $28.1^\circ$
for $\alpha=0$.  The difference in the orbital inclinations with
respect to the sky plane $|i_1-i_2|$ is a lower limit to the true
mutual inclination, and thus under these assumptions the orbits are
misaligned at the $>95\%$ c.l. 

\begin{figure}
\includegraphics[scale=0.7]{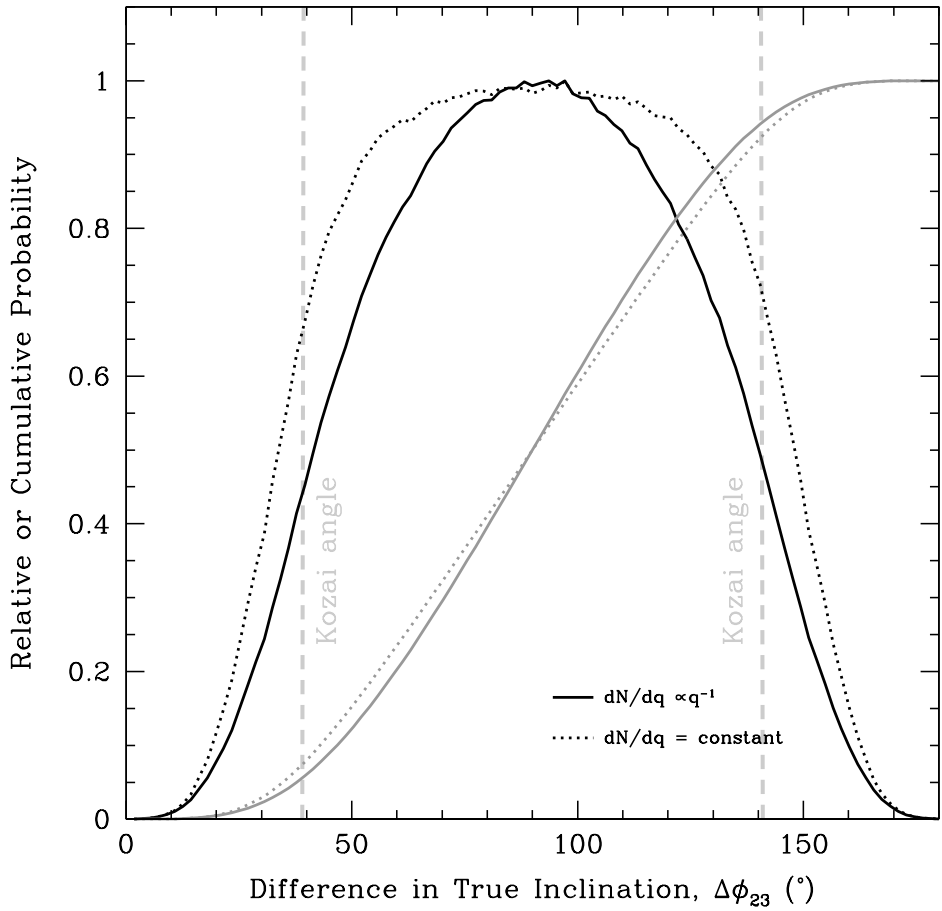}
\caption{The a posteriori probability density (black lines) and
cumulative probability (grey lines) for the true relative inclination
of the secondary and tertiary $\Delta\phi_{23}$.  The probability densities
have been arbitrarily normalized by their peak probability density.
The solid lines show the probabilities assuming a prior on the mass
ratio $q$ of $dN/dq \propto q^{-1}$ (uniform in $\log{q}$), where as
the dotted lines show $dN/dq$ = constant (uniform in $q$).  In both
cases, tidal synchronization/coplanarity for the secondary has been
assumed, and companion masses $\ge 1 ~ \rm{M_\odot}$ have been excluded.  The
curves are symmetric about $\Delta \phi=90^\circ$ because translating
the inclinations of the secondary and tertiary by $\pi$ results in the
same observables.
\label{dib}}
\end{figure}

The true mutual inclination $\Delta \phi_{23}$ of the orbits is
related to the inclinations referenced to the sky plane by,
\begin{equation}
\cos \Delta \phi_{23} = \cos i_2 \cos i_3 + \sin i_2 \sin i_3
\cos(\Delta\Omega_{23}),
\label{eqn:cosphi}
\end{equation}
where $\Delta\Omega_{23}\equiv \Omega_2-\Omega_3$ and $\Omega_j$ is
the longitude of the ascending node for companion $j$. Fig.\ \ref{dib}
shows the distribution of mutual inclinations assuming the flux
ratio constraint, a uniform distribution for
$\Delta\Omega_{12}$, and the two different priors on $\alpha$. These
distributions are symmetric about $\Delta \phi_{12}=90^\circ$ because
the observables (radial velocity amplitude and potential transits) are
invariant under the transformations $i_j\rightarrow i_j+\pi$.  
The $95\%$ c.l. lower and upper limits on the true mutual inclinations
are $31.1{^\circ}$ and $180{^\circ}-31.1{^\circ} = 148.9{^\circ}$ for $\alpha=-1$ and $28.8{^\circ}$ and $180{^\circ}-28.8{^\circ}
= 151.2{^\circ}$ for $\alpha=0$. The probability that the mutual inclination
is greater than the Kozai angle \citep{koz1962} of $39.2{^\circ}$ or
less than the retrograde Kozai angle of $140.8{^\circ}$ is $\sim 89\%$
($\alpha=-1$) and $\sim 85\%$ ($\alpha=0$).

\begin{figure}
\includegraphics[scale=0.7]{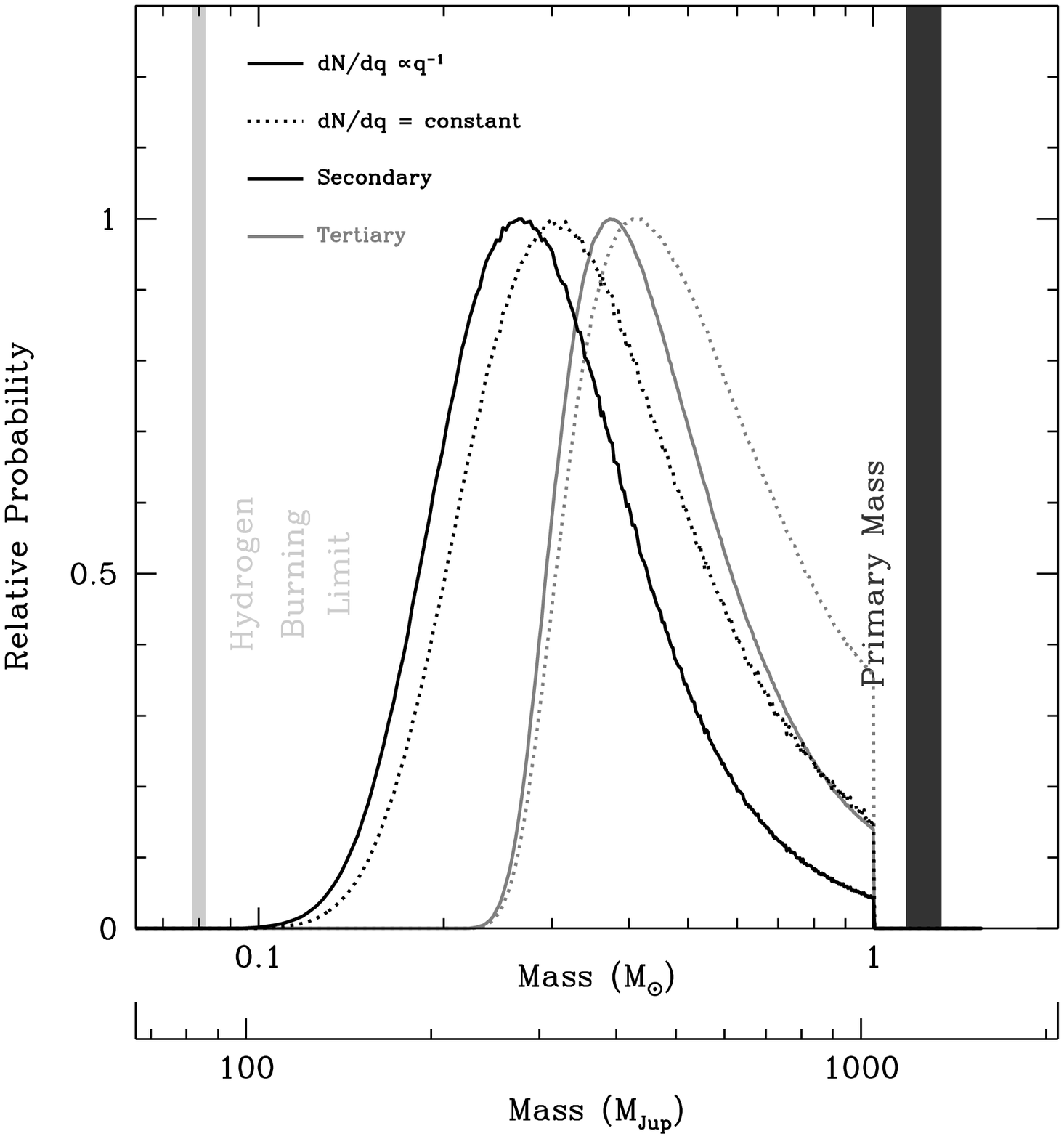}
\caption{The a posteriori probability density for the true masses of
the tertiary and secondary.  These have been arbitrarily normalized by
their peak probability density.  The black lines show the
probabilities for the secondary assuming the tidal
synchronization/coplanarity, for two different priors on the mass
ratio $q$, $dN/dq \propto q^{-1}$ (uniform in $\log{q}$, solid), and
$dN/dq$ = constant (uniform in $q$, dotted).  The grey lines show
the probabilities for the tertiary, for the same priors on the mass
ratio.  In all cases,
companion masses $\ge 1 ~ \rm{M_\odot}$ have been excluded.  
\label{mb}}
\end{figure}

Fig.\ \ref{mb} shows the probability densities of the true mass of the
secondary and tertiary, assuming the tidal synchronization/coplanarity
and flux ratio constraints, for $\alpha=-1$ and $\alpha=0$. We infer
that the primary and secondary are both likely to be M dwarfs, with
the tertiary likely to be somewhat more massive than the secondary.

\section{The Kozai-Lidov Mechanism Applied To TYC 2930}
\label{kozaisection}
One mechanism of forming short-period binary stars is the Kozai-Lidov mechanism, in which a hierarchical triple system exchanges angular momentum between the inner and outer orbits periodically if the mutual inclination between the inner and outer orbits is $39.2^{\circ} \lesssim {\Delta \phi_{23}} \lesssim 141.8^{\circ}$ \citep{koz1962,lid1962}.  These oscillations drive $e_2$ to high and low values while decreasing and increasing $\Delta \phi_{23}$, respectively.  When combined with tidal friction, these oscillations can cause the semimajor axis of the inner companion ($a_{2}$) to decrease, as tidal friction removes energy and circularizes the orbit \citep{maz1979}.  In addition to short-period binaries in triple systems \citep{kis1998, egg2001, tok2006, fab2007}, this mechanism has been proposed as an explanation for the prevalence of ``Hot Jupiter'' gas giant planets with periods of a few days \citep{wu2003,fab2007,wu2007,nao2011} and the formation of blue straggler stars in globular clusters \citep{per2009}.

To constrain the initial parameters of TYC 2930's progenitor system, we first assume that $a_3$ and $e_3$ remain unchanged during the evolution of the system.  We can then place an upper limit on the initial $a_2$ ($a_{2,\rm{i}}$) using the stability criteria for triple systems given by \citep{mar2001}:
\begin{equation}
  \frac{a_3}{a_2} \geq  C \: f \left[ \left( 1 + \frac{M_3}{M_1 + M_2} \right) \frac{1 + e_3}{(1 - e_3)^{3}} \right] ^{0.4},
 \label{eq:mardling}
\end{equation}
where $C=2.8$ is an empirically-fit constant and $f=1-\frac{0.3}{\pi}\:{\Delta \phi_{23}}$ is an ad-hoc mutual inclination term.  Given the most probable values for the parameters of TYC 2930, and assuming that the initial ${\Delta \phi_{23}} \sim 90^{\circ}$ such that the Kozai-Lidov mechanism is a significant effect, the initial $a_3/a_2 > 4.4$, and thus the maximum stable $a_2$ is $a_{2\rm{,max}} = 0.96$ AU.  To replicate the observed system parameters, we then require $r_{\rm{peri}} < 3 \, R_1 = r_{\rm{tide}}$ for efficient tidal circularization.  Generally, the $\cos{\Delta \phi_{23}}$ required for $r_{\rm{peri}} < 3 \; R_1$, $\cos{\Delta \phi_{23}}_{\rm{crit}}$, is given by:
\begin{equation}
 \cos{\Delta \phi_{23}}_{\rm{crit}} = \left( \frac{3}{5} \left[1 - \left( 1 - \frac{r_{\rm{tide}}}{a_{2\rm{,i}}} \right)^2 \right] \right)^{1/2} \sim \left( \frac{6\, r_{\rm{tide}}}{5\, a_{2\rm{,i}}} \right)^{1/2},
 \label{eq:cosicrit}
\end{equation}
Thus if $a_{2\rm{,i}} = 0.3$ AU then $\cos{\Delta \phi_{23}}_{\rm{crit}} \sim 0.25$.  

\begin{figure}[t]
  \centerline{
    \includegraphics[width=7.0cm, angle=90]{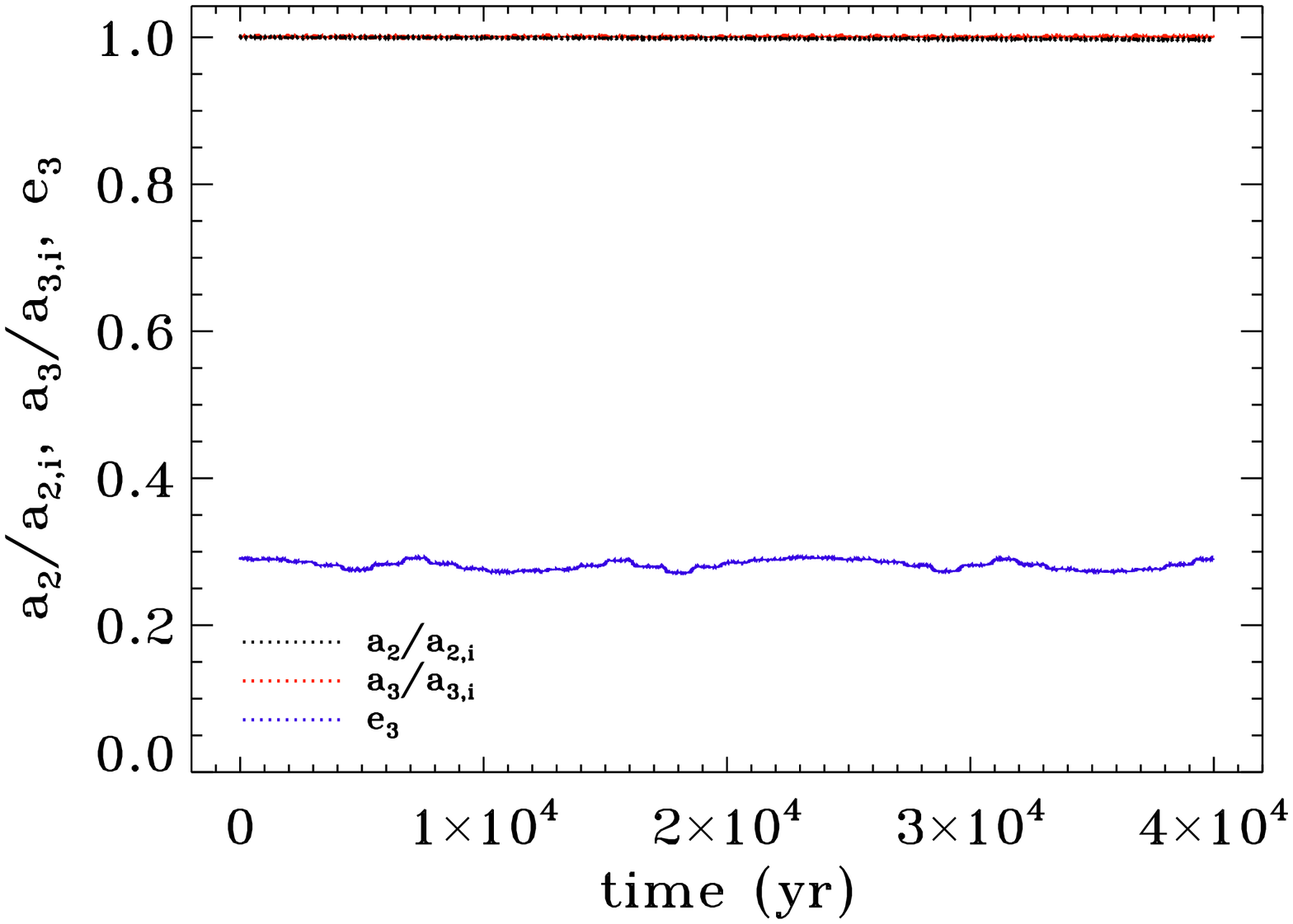}
  }
  \centerline{
    \includegraphics[width=7.0cm, angle=90]{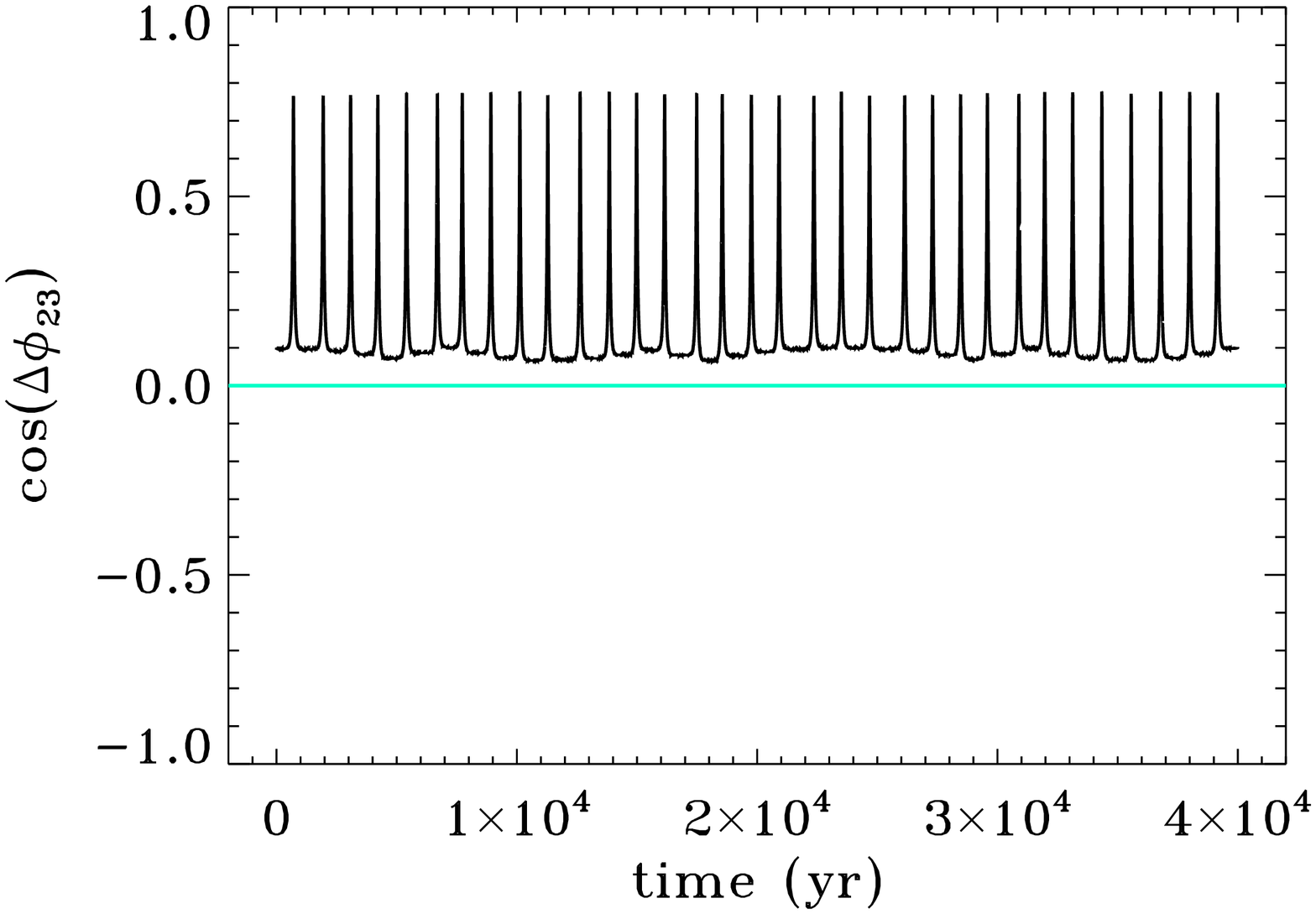}
  }
  \centerline{
    \includegraphics[width=7.0cm, angle=90]{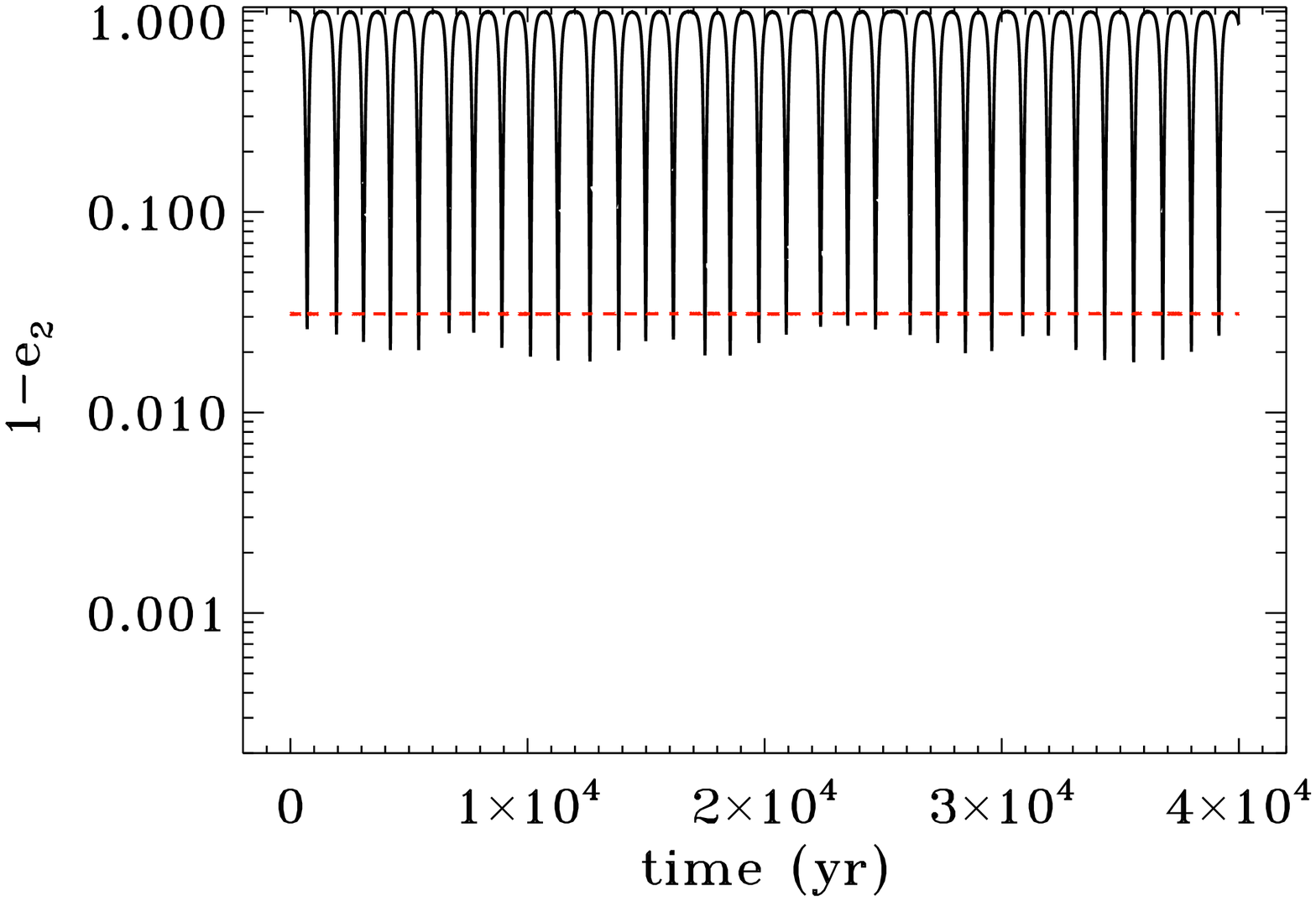}
  }
  \caption{Example of the normal Kozai-Lidov mechanism using parameters consistent with those of TYC 2930's progenitor system. {\it Top:} Evolution of $a_2$ (dotted black), $a_3$ (red dotted), and $e_3$ (blue dotted).  {\it Middle:} Time evolution of $\cos{\Delta \phi_{23}}$.  The teal line designates $\cos{\Delta \phi_{23}}=0$. {\it Bottom:} Time evolution of $e_2$, the red line denotes $r_{\rm{peri}} < 2 ~ \rm{R_{\odot}}$. Integrations after the first eccentricity maximum are for illustration purposes only, since we do not include the effects of tides here.}
\label{normalkozai}
\end{figure}

To simulate the Kozai-Lidov mechanism, we use a modified version of the $N$-body code FEWBODY\footnote{FEWBODY is now available at \url{http://fewbody.sourceforge.net/}.} \citep{fre2004}.  In Fig.\ \ref{normalkozai}, we show an example triple system which has initial parameters consistent with the constraints placed on the progenitor system of TYC 2930: $M_1 = 1.21 ~ \rm{M_{\odot}}$, $M_2 = 0.34 ~ \rm{M_{\odot}}$, $M_3 = 0.48 ~ \rm{M_{\odot}}$, $a_{2\rm{,i}} = 0.3 ~ \rm{AU}$, $a_3 = 4.27 ~ \rm{AU}$, $e_2 = 0.0$, $e_3 = 0.29$, $\omega_2 = 0^{\circ}$, $\omega_3 = 0^{\circ}$, and $\cos{\Delta \phi_{23}} = 0.1$.  The top panels present the evolution of $a_2$, $a_3$ and $e_3$.  The middle and bottom panels show the evolution of $\cos{\Delta \phi_{23}}$ and $1 - e_2$, respectively.  In the middle panel, the teal line designates $\cos{\Delta \phi_{23}} = 0$.  In the bottom panel, the red line shows where $e_2$ is high enough such that the radius of periastron of the inner binary ($r_{\rm{peri}}$) is equal to $2 ~ \rm{R_{\odot}}$.  The example system goes through Kozai-Lidov cycles, bringing $r_{\rm{peri}} < 2 ~ \rm{R_{\odot}}$.  If the effects of tides were included in our calculation, we would expect that $a_2$ and ${\Delta \phi_{23}}$ would decrease over many Kozai cycles.

However, the commonly-used, quadrupole-order expansion of the three-body Hamiltonian is insufficient to capture the secular dynamics of triple systems under the test particle approximation when $e_3$ is non-zero \citep{lit2011, kat2011, nao2011}. The importance of the octupole-order terms relative to the quadrupole-order terms in the doubly-averaged three-body Hamiltonian is given by the parameter
\begin{equation}
 \epsilon_{\rm{oct}} = \left( \frac{M_0 - M_1}{M_0 + M_1} \right) \left(\frac{a_2}{a_3} \right) \frac{e_3}{1 - e_3^{2}}.
 \label{eq:epsoct}
\end{equation}
In the limit that $(M_2 \ll M_1, M_3)$ and $e_3 \neq 0$, it is possible for the triple system to ``flip'', i.e., the system exhibits quasi-periodic cycles in $\Delta \phi_{23}$ through 0, and the tertiary passes between prograde and retrograde \citep{lit2011, kat2011}.  These flips occur even for small values of $\epsilon_{\rm{oct}}$ ($\sim 10^{-3}$) in the test particle approximation, as long as the system is sufficiently inclined and the arguments of periastrons are chosen judiciously (see Figures 3 and 7 of \citealp{kat2011} and \citealp{lit2011}, respectively).  

Flips become increasingly common for triple systems with larger values of $\epsilon_{\rm{oct}}$, and they correspond to extremely large spikes in $e_2$, such that $\left(1-e_2\right) \sim 10^{-5}$.  Such qualitative behaviors (flips and eccentricity spikes) are referred to as the ``eccentric Kozai-Lidov mechanism'', and can occur over a broad range of parameters within octupole-order calculations.  \citet{sha2012} have recently highlighted the mass dependence of the eccentric Kozai-Lidov mechanism, and performed one of the first explorations of the eccentric Kozai-Lidov mechanism for triple stellar systems.  They found that many triple systems become tidally affected while on the main sequence as a result of this mechanism. The mechanism can even be effective for triple systems that would not otherwise become tidally affected during normal Kozai-Lidov oscillations in quadrupole-order calculations.

\begin{figure}[t]
  \centerline{
    \includegraphics[width=7.0cm, angle=90]{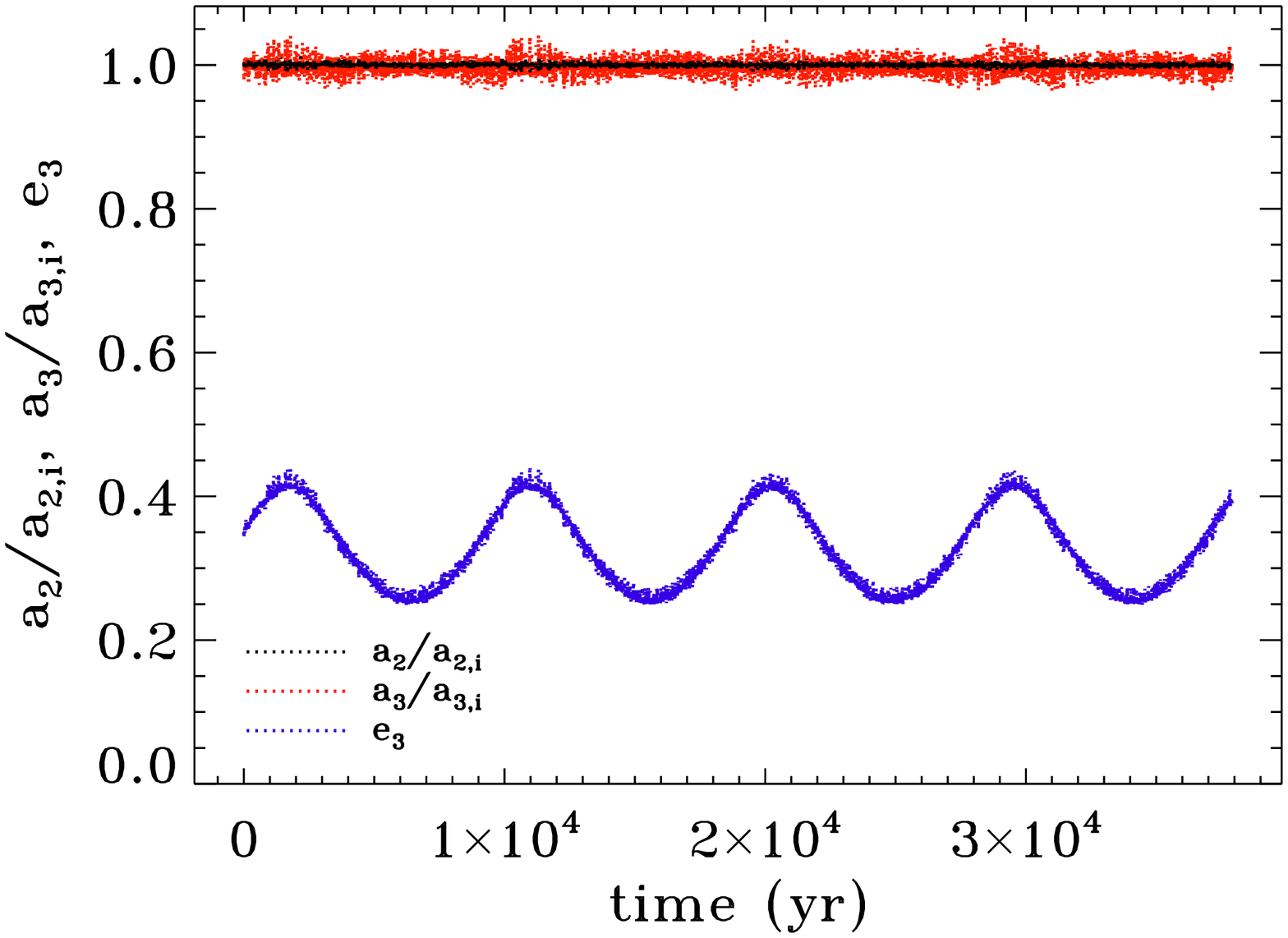}
  }
  \centerline{
    \includegraphics[width=7.0cm, angle=90]{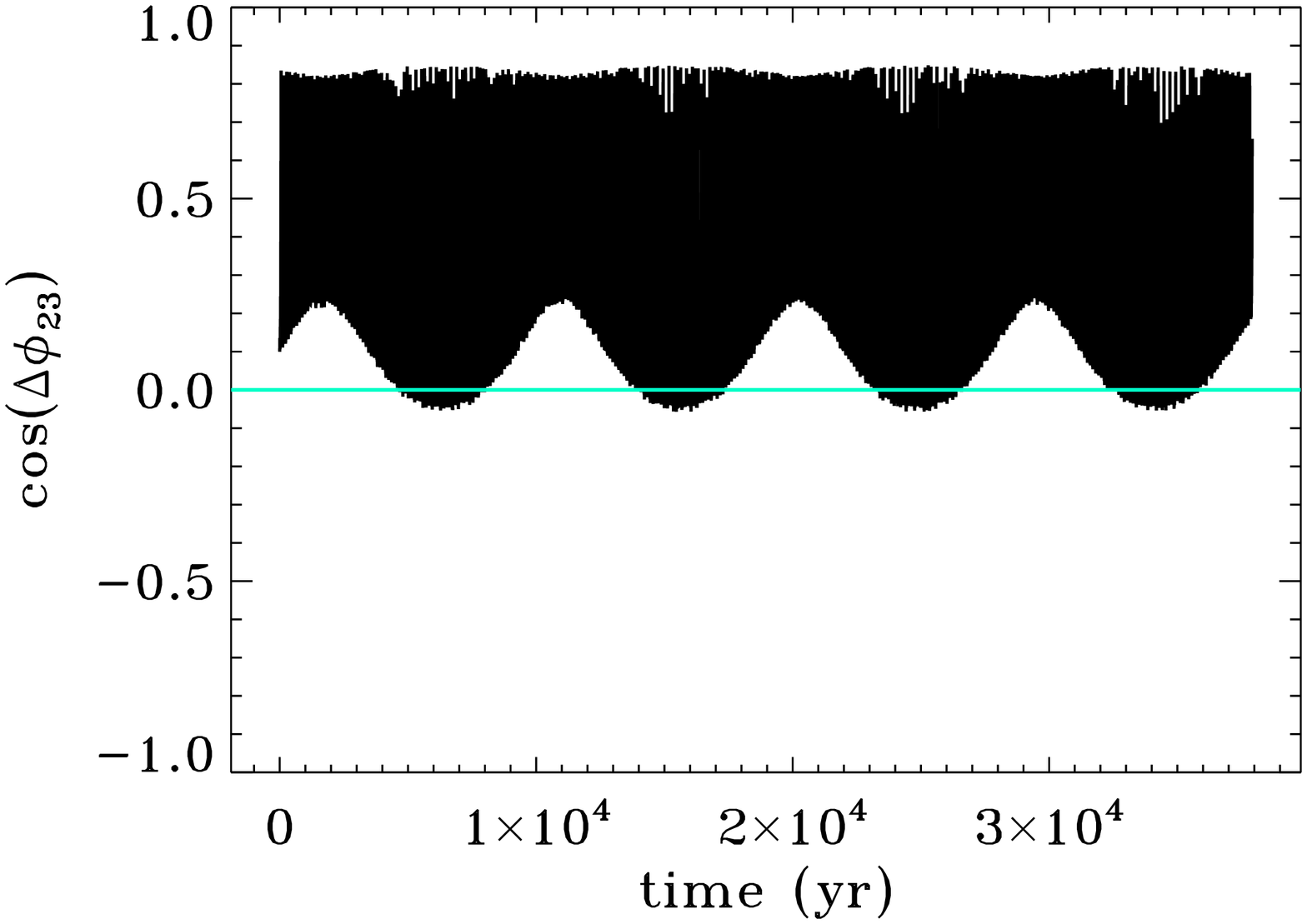}
  }
  \centerline{
    \includegraphics[width=7.0cm, angle=90]{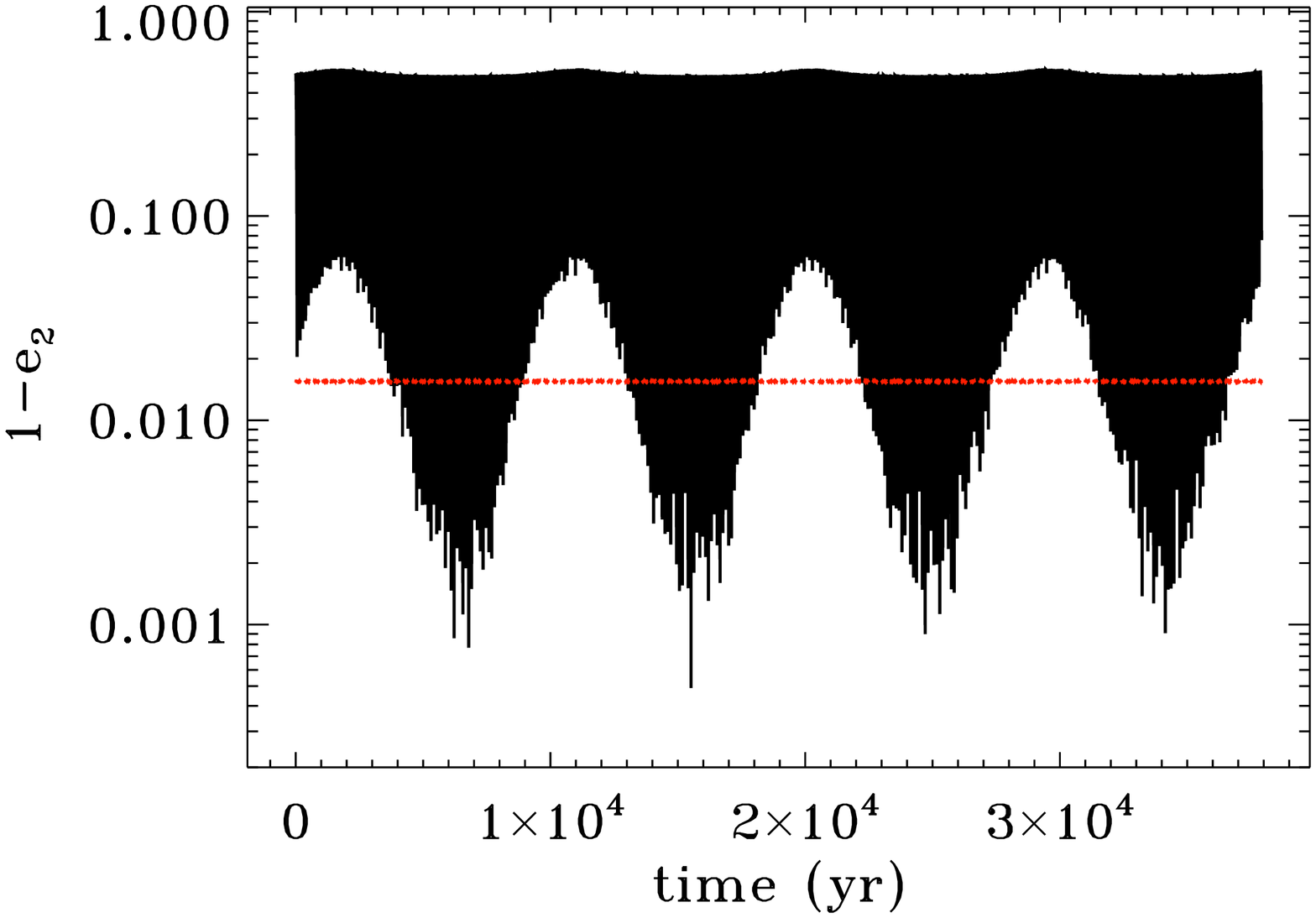}
  }
  \caption{Similar to Fig.\ \ref{normalkozai}, but for the eccentric Kozai-Lidov mechanism.  Note the qualitative differences between the two cases: the changing sign of $\cos{\Delta \phi_{23}}$ and the extreme eccentricity spikes.}
  \label{eccentrickozai}
\end{figure}

As an example of the eccentric Kozai-Lidov mechanism, we integrated the orbit of a system with $M_1 = 1.21 ~ \rm{M_{\odot}}$, $M_2 = 0.34 ~ \rm{M_{\odot}}$, $M_3 = 0.48 ~ \rm{M_{\odot}}$, $a_2 = 0.6 ~ \rm{AU}$, $a_3 = 4.27 ~ \rm{AU}$, $e_2 = 0.5$, $e_3 = 0.35$, $\omega_2 = 0^{\circ}$, $\omega_3 = 262^{\circ}$, and $\cos{\Delta \phi_{23}} = 0.1$. Fig.\ \ref{eccentrickozai} shows the resulting evolution.  There are many qualitative differences between the evolution of this system and that shown in Fig.\ \ref{normalkozai}.  The most obvious of these are the periodic oscillations of $e_3$, the long-term oscillations of the minimum $\cos{\Delta \phi_{23}}$, and the flipping of the sign of $\cos{\Delta \phi_{23}}$, which corresponds to large spikes in $e_2 \sim 0.9995$.

What is the fate of systems that become tidally affected due to the eccentric Kozai-Lidov mechanism?  Unfortunately, no study has been performed to investigate the eccentric Kozai-Lidov mechanism with the inclusion of tidal dissipation for triple stellar systems.  It is possible that tidal friction will simply detune the eccentric Kozai-Lidov mechanism, inhibiting the extreme spikes in eccentricity observed in our example system. However, \citet{nao2011} investigated the eccentric Kozai-Lidov mechanism with tidal friction for a Jupiter-mass secondary in an octupole-order calculation.  They demonstrate that the extremely high $e_2$ obtained during a flip can lead to large tidal effects, which rapidly capture the planet in a short-period retrograde orbit.  They coin this mechanism ``Kozai capture'' and claim it may explain the occurrence of retrograde, short-period, gas giant planets.  A similar mechanism may operate for stellar triple systems, in which Kozai capture would cause a rapid dissipation of orbital energy from the inner binary during its evolution.  This would result in a circularized inner binary with a semimajor axis of a few $r_{\rm{peri}}$, and thus provides a plausible scenario by which TYC 2930's secondary star was driven to such a small semimajor axis.

\section{Conclusion}
We have discovered a short-period companion to TYC 2930-00872-1 with a
minimum mass below the hydrogen burning limit.  Despite its relatively high
transit probability, we exclude any transits of the companion with
high confidence using data from three ground-based telescopes.  A
long-term trend in the RVs indicate the presence of a longer-period
tertiary in the system.  The tertiary's spectral lines are not detected in 
our spectroscopic data, its fluxes do not significantly contribute to our SED fitting, and direct imaging excludes
stellar-mass, main-sequence companions out to projected separations of
30 AU.  Our spectra show the clear presence of Ca H and K core
emission, but there is an unexpected lack of photometric variability, and
the measured $v\sin I$ is significantly smaller than expected if the
primary's rotation rate was tidally synchronized to the inner
companion's orbital period.  This suggests the primary's stellar spin
axis is closely aligned to the line-of-sight.  Given the age
of the system, it is expected that the inner companion's orbital
angular momentum vector is aligned with the stellar spin axis,
therefore its line-of-sight orbital inclination is low, and its true
mass is likely to be stellar.  The absence of any detected signal from either component in the spectra
and SED place an upper mass limit of $\sim 1.0 ~ \rm{M_{\odot}}$, if they are not stellar remnants.  Assuming the tertiary
is not a remnant, the upper mass limit places a lower limit to its
line-of-sight inclination, which results in a significant mutual
inclination between the secondary and tertiary.  Such mutual
inclinations are expected if the system's dynamical history was driven
by the Kozai-Lidov mechanism.  Long-term RV monitoring of the outer
companion to obtain reliable orbital parameters will greatly improve the
constraints that can be placed on the mutual inclination between the
secondary and tertiary.  Furthermore, high signal-to-noise ratio 
spectroscopic observations that could detect the presence of the
(presumably) M dwarf companions would also allow for masses to be
assigned to both objects and improve the inclination constraints.

\acknowledgements
Keivan Stassun, Leslie Hebb, and Joshua Pepper acknowledge funding support from the Vanderbilt Initiative in Data-Intensive Astrophysics (VIDA) from Vanderbilt University, and from NSF Career award AST-0349075.  Eric Agol thanks the NSF for Career grant AST-0645416.  John Wisniewski acknowledges support from NSF Astronomy \& Astrophysics Postdoctoral Fellowship AST 08-02230.  Luan Ghezzi acknowledges financial support provided by the PAPDRJ CAPES/FAPERJ Fellowship.  Leticia Dutra-Ferreira acknowledges financial support provided by CAPES and ESO student fellowship.  Gustavo F. Porto de Mello acknowledges financial support from CNPq grants no. 476909/2006-6 and 474972/2009-7, plus a FAPERJ grant no. APQ1/26/170.687/2004.  Chelsea Vincent and Gwendolyn Weaver acknowledge support from the Pennsylvania Space Grant Consortium.  Operation of Allegheny Observatory is supported in part by the Theiss Memorial Endowment.  Work by B.\ Scott Gaudi and Thomas Beatty was partially supported by NSF Career grant AST-1056524.  We thank J. Fregeau for making the code FEWBODY publicly available. This work is supported in part by an Alfred P. Sloan Foundation Fellowship and NSF grant AST-0908816.  Benjamin Shappee was supported by a Graduate Research Fellowship from the National Science Foundation.

This work was partially supported by funding from the Center for Exoplanets and Habitable Worlds. The Center for Exoplanets and Habitable Worlds is supported by the Pennsylvania State University, the Eberly College of Science, and the Pennsylvania Space Grant Consortium.  This research has made use of the SIMBAD database, operated at CDS, Strasbourg, France.  This publication makes use of data products from the Two Micron All Sky Survey, which is a joint project of the University of Massachusetts and the Infrared Processing and Analysis Center/California Institute of Technology, funded by the National Aeronautics and Space Administration and the National Science Foundation.  This publication makes use of data products from the Wide-field Infrared Survey Explorer, which is a joint project of the University of California, Los Angeles, and the Jet Propulsion Laboratory/California Institute of Technology, funded by the National Aeronautics and Space Administration.

Funding for the MARVELS multi-object Doppler instrument was provided by the W.M. Keck Foundation and NSF with grant AST-0705139.  The MARVELS survey was partially funded by the SDSS-III consortium, NSF grant AST-0705139, NASA with grant NNX07AP14G and the University of Florida.  This work has made use of observations taken with the Telescopio Nationale Galileo (TNG) operated on the island of La Palma by the Fundation Galileo Galilei, funded by the Instituto Nazionale di Astrofisica (INAF), in the Spanish {\it Observatorio del Roque de los Muchachos} of the Instituto de Astrof{\'\i}sica de Canarias (IAC). 

This work was based on observations with the SDSS 2.5-meter telescope.  Funding for SDSS-III has been provided by the Alfred P. Sloan Foundation, the Participating Institutions, the National Science Foundation, and the U.S. Department of Energy Office of Science. The SDSS-III web site is \url{http://www.sdss3.org/}.  SDSS-III is managed by the Astrophysical Research Consortium for the Participating Institutions of the SDSS-III Collaboration including the University of Arizona, the Brazilian Participation Group, Brookhaven National Laboratory, University of Cambridge, Carnegie Mellon University, University of Florida, the French Participation Group, the German Participation Group, Harvard University, the Instituto de Astrofisica de Canarias, the Michigan State/Notre Dame/JINA Participation Group, Johns Hopkins University, Lawrence Berkeley National Laboratory, Max Planck Institute for Astrophysics, Max Planck Institute for Extraterrestrial Physics, New Mexico State University, New York University, Ohio State University, Pennsylvania State University, University of Portsmouth, Princeton University, the Spanish Participation Group, University of Tokyo, University of Utah, Vanderbilt University, University of Virginia, University of Washington, and Yale University.


\clearpage

\begin{deluxetable}{rrrr}
\tabletypesize{\scriptsize}
\tablecaption{MARVELS RV - TYC 2930\label{marvrvs}}
\tablewidth{0pt}
\tablehead{
\colhead{HJD$_{\rm{UTC}}$} & \colhead{RV} & \colhead{$QF$-Scaled $\sigma_{\rm{RV}}$}\\
\colhead{~} & \colhead{(km s$^{-1}$)} & \colhead{(km s$^{-1}$)}
}
\startdata
2454843.86946 &   2.571 &  0.129  \\
2454844.82593 &  -8.036 &  0.099  \\
2454845.83333 &  -6.326 &  0.095  \\
2454846.82454 &   1.039 &  0.077  \\
2454847.77335 & -13.549 &  0.083  \\
2454866.71528 &  -7.996 &  0.087  \\
2454874.74327 & -10.879 &  0.118  \\
2454876.77555 & -12.778 &  0.128  \\
2455135.87308 &   9.021 &  0.120  \\
2455136.84418 &  -7.602 &  0.105  \\
2455137.85985 &   7.069 &  0.165  \\
2455138.88972 &  -1.846 &  0.120  \\
2455139.78368 &  -3.523 &  0.110  \\
2455139.96991 &   0.506 &  0.099  \\
2455143.86472 &  -4.111 &  0.102  \\
2455144.86232 &   1.319 &  0.084  \\
2455145.86470 &   5.141 &  0.107  \\
2455171.89913 &   7.588 &  0.094  \\
2455172.85712 &  -0.125 &  0.118  \\
2455200.91808 &   5.337 &  0.077  \\
2455254.77584 &  10.438 &  0.085  \\
2455258.75174 &  -4.726 &  0.104  \\
2455281.65360 &  10.108 &  0.141  \\
2455466.86921 &   1.955 &  0.137  \\
2455466.91135 &   0.996 &  0.105  \\
2455487.84516 &   7.664 &  0.077  \\
2455488.86530 &  -0.942 &  0.101  \\
2455489.87545 &  -0.646 &  0.095  \\
2455494.90526 &   3.061 &  0.099  \\
2455500.90889 &   1.409 &  0.092  \\
2455521.82966 &   6.852 &  0.107  \\
2455522.85494 &  -0.199 &  0.117  \\
2455550.88590 &   4.448 &  0.195  \\
\enddata
\end{deluxetable}

\begin{deluxetable}{rrr}
\tabletypesize{\scriptsize}
\tablecaption{SARG RV - TYC 2930\label{tngrvs}}
\tablewidth{0pt}
\tablehead{
\colhead{HJD$_{\rm{UTC}}$} & \colhead{RV} & \colhead{$\sigma_{\rm{RV}}$} \\
\colhead{~} & \colhead{(km s$^{-1}$)} & \colhead{(km s$^{-1}$)}
}
\startdata
2455436.71196 &   2.568 & 0.026 \\
2455460.73287 &  -3.624 & 0.015 \\
2455460.74384 &  -3.400 & 0.018 \\
2455460.75505 &  -3.196 & 0.021 \\
2455495.57813 &   5.964 & 0.034 \\
2455495.61314 &   5.550 & 0.020 \\
2455495.69113 &   4.642 & 0.024 \\
2455495.71406 &   4.247 & 0.023 \\
2455516.55620 &  -5.453 & 0.022 \\
2455516.63469 &  -3.758 & 0.012 \\
2455553.50056 &   4.252 & 0.014 \\
2455553.66834 &   5.920 & 0.013 \\
2455580.42134 &   5.869 & 0.012 \\
2455580.46981 &   6.010 & 0.012 \\
2455580.59158 &   5.770 & 0.014 \\
2455666.42077 &  -8.355 & 0.014 \\
2455698.35676 & -12.133 & 0.016 \\
2455791.70794 &   3.088 & 0.012 \\
2455844.61269 &  -8.335 & 0.015 \\
2455844.74543 &  -5.583 & 0.012 \\
\enddata
\end{deluxetable}

\begin{deluxetable}{lrr}
\tabletypesize{\scriptsize}
\tablecaption{Stellar Properties - TYC 2930\label{stellarprops}}
\tablewidth{0pt}
\tablehead{
\colhead{Parameter} & \colhead{Value} & \colhead{$\pm 1{\sigma}$}
}
\startdata
$\alpha_{\rm{J2000}}$ (deg)\tablenotemark{a} &  93.880921  & 0.000004 \\
$\delta_{\rm{J2000}}$ (deg)\tablenotemark{a} & +39.931826  & 0.000005 \\
$FUV$\tablenotemark{b} & 19.815 & 0.195 \\
$NUV$\tablenotemark{b} & 14.34 & 0.01 \\
$B$ (HAO)   & 10.365 & 0.023 \\
$V$ (HAO)   & 9.842 & 0.018 \\
$J$ (2MASS) & 8.770 & 0.029 \\
$H$ (2MASS) & 8.539 & 0.047 \\
$K_{S}$ (2MASS) & 8.458 & 0.023 \\
$WISE ~ 3.4 ~ \rm{{\mu}m}$ & 8.380 & 0.024 \\
$WISE ~ 4.6 ~ \rm{{\mu}m}$ & 8.392 & 0.023 \\
$WISE ~ 12 ~ \rm{{\mu}m}$  & 8.329 & 0.029 \\
$WISE ~ 22 ~ \rm{{\mu}m}$  & 8.201 & 0.245 \\
$\mu_{\alpha} \left(\rm{mas ~ yr^{-1}}\right)$\tablenotemark{c} & 3.42 & 2.05 \\
$\mu_{\delta} \left(\rm{mas ~ yr^{-1}}\right)$\tablenotemark{c} & -46.69 & 1.13 \\
Parallax $\Pi \left(\rm{mas}\right)$\tablenotemark{c} & 7.15 & 1.51 \\[1ex]
$A_{\rm{V}}$ (SED) & 0.33 & 0.06 \\
$T_{\rm eff} ~ \left(\rm{K}\right)$ & 6427 & 33 \\
$\log{(g ~ {\rm [cm ~ s^{-1}]})}$ & 4.52 & 0.14 \\
$\rm{[Fe/H]}$ & -0.04 & 0.05 \\
${\xi}_{t} ~ \left(\rm{km ~ s^{-1}}\right)$ & 1.40 & 0.05 \\[1ex]
$v\sin I ~ \left(\rm{km ~ s^{-1}}\right)$ & 3.8 & $^{+1.9} _{-2.8}$ \\
$M_{*}$ ($\rm{M_{\odot}}$) & 1.21 & 0.08 \\[1ex]
$R_{*}$ ($\rm{R_{\odot}}$) & 1.09 & $^{+0.15} _{-0.13}$ \\
$\rm{RPM_{J}}$ & 2.14 & - \\
\enddata
\tablenotetext{a}{Tycho-2 Catalog \citep{hog2000}}
\tablenotetext{b}{GALEX \citep{mar2005}}
\tablenotetext{c}{\citet{van2007}}
\end{deluxetable}

\begin{deluxetable}{lllll}
\tabletypesize{\scriptsize}
\tablecaption{IAC and BPG Stellar Parameters \label{bpgiacsp}}
\tablewidth{0pt}
\tablehead{
\colhead{} & \multicolumn{2}{c}{IAC} & \multicolumn{2}{c}{BPG}\\
\colhead{Parameter} & \colhead{SARG} & \colhead{ARCES} & \colhead{SARG} & \colhead{ARCES}
}
\startdata
\ion{Fe}{1} lines used & 173 & 172 & 60 & 67 \\
\ion{Fe}{2} lines used & 21 & 25 & 8 & 9 \\
$T_{\rm eff} ~ \left(\rm{K}\right)$ & $6456 \pm 49$ & $6413 \pm 41$ & $6406 \pm 110$ & $6415 \pm 76$ \\
$\log{(g ~ {\rm [cm ~ s^{-1}]})}$ & $4.68 \pm 0.27$ & $4.53 \pm 0.21$ & $4.47 \pm 0.26$ & $4.44 \pm 0.21$ \\
$\rm{[Fe/H]}$ & $-0.02 \pm 0.09$ & $-0.01 \pm 0.07$ & $-0.13 \pm 0.10$ & $-0.03 \pm 0.07$ \\
${\xi}_{t} ~ \left(\rm{km ~ s^{-1}}\right)$ & $1.296 \pm 0.076$ & $1.464 \pm 0.059$ & $1.44 \pm 0.20$ & $1.34 \pm 0.12$ \\
\enddata
\end{deluxetable}

\begin{deluxetable}{rrrr}
\tabletypesize{\scriptsize}
\tablecaption{Orbital Parameters For The Inner Companion\label{orbprops}}
\tablewidth{0pt}
\tablehead{
\colhead{Parameter} & \colhead{Median} & \colhead{$\sigma$ (Low)} & \colhead{$\sigma$ (High)}
}
\startdata
$P$ (days) & 2.430420 & 0.000006 & 0.000006 \\
$K$ (km s$^{-1}$) & 8.723 & 0.009 & 0.009 \\
$T_c \left({\rm{HJD_{UTC} - 2450000.0}}\right)$ & 4842.2640 & 0.00187 & 0.00187 \\
$e$ & 0.0066 & 0.0010 & 0.0010 \\
141.948
$\omega$ & 142 & 7 & 7 \\
$\gamma_{0,\rm{inst}} ~ \left(\rm{km ~ s^{-1}}\right)$ & -6.642 & 2.106 & 0.904 \\
$\gamma_{\rm{off}} ~ \left(\rm{km ~ s^{-1}}\right)$ & -2.890 & 2.102 & 0.900 \\
$\gamma_{0} ~ \left(\rm{km ~ s^{-1}}\right)$ & 35.751 & 0.285 & 0.285 \\
\enddata
\end{deluxetable}

\begin{deluxetable}{rrrrr}
\tabletypesize{\scriptsize}
\tablecaption{A priori secondary mass. \label{apriorimass1}}
\tablewidth{0pt}
\tablehead{
\colhead{$\alpha$\tablenotemark{a}} & \colhead{key\tablenotemark{b}} & \colhead{Median $M_2$, 68\%} & \colhead{Median $i_2$, 68\%} & \colhead{Transit Prob.}
}
\startdata
-1 &	1 &	$0.080_{-0.012}^{+0.048}$ & $56.0_{-24.8}^{+18.7}$ & 0.133 \\[1ex]
0  &	1 &	$0.098_{-0.028}^{+0.160}$ & $42.4_{-26.3}^{+27.2}$ & 0.086 \\[1ex]
-1 &	2 &	$0.338_{-0.110}^{+0.216}$ & $12.6_{-4.2}^{+5.2}$   & $<0.001$ \\[1ex]
0  &	2 &	$0.416_{-0.150}^{+0.280}$ & $10.6_{-3.6}^{+5.0}$   & $<0.001$ \\[1ex]
\enddata
\tablenotetext{a }{Prior of the form: $N/dq \propto q^{\alpha}$}
\tablenotetext{b }{1=Flux Ratio \& Transit Constraint. 2=Flux Ratio, Transit, and Synchronization/Coplanarity.}
\end{deluxetable}

\begin{deluxetable}{rrrr}
\tabletypesize{\scriptsize}
\tablecaption{A priori tertiary mass. \label{apriorimass2}}
\tablewidth{0pt}
\tablehead{
\colhead{$\alpha$} & \colhead{Median $M_3$, 68\%} & \colhead{Median $i_3$, 68\%} & \colhead{Transit Prob.}
}
\startdata
-1 &	$0.480_{-0.131}^{+0.245}$ &  $65.4_{-21.2}^{+17.1}$ &  0.001 \\[1ex]
0  &	$0.553_{-0.173}^{+0.261}$ &  $61.5_{-25.4}^{+19.5}$ &  0.001 \\[1ex]
\enddata
\tablenotetext{a }{Includes Flux Ratio Constraint}
\end{deluxetable}


\clearpage

\clearpage

\clearpage

\end{document}